\documentclass[twocolumn,prx,showpacs,preprintnumbers,amsmath,amssymb,nofootinbib]{revtex4}

\usepackage[utf8]{inputenc} 
\usepackage[T1]{fontenc}    
\usepackage{bm}
\usepackage{color}
\usepackage{xcolor}
\usepackage{empheq}
\usepackage{amsfonts}       
\usepackage{nicefrac}     
\usepackage{url}  
\usepackage{amssymb}
\usepackage{algorithm}
\usepackage{algorithmic}

\usepackage{paperstyle}        

\begin{document}

\title{Phase transitions in the $q$-coloring of random hypergraphs}

\author{Marylou Gabri\'e}
\affiliation{Laboratoire de physique statistique, D\'epartement de physique de l'ENS, Ecole normale sup\'erieure, Universit\'e Paris Diderot, UPMC Univ. Paris 06, CNRS, PSL Research University, 75005 Paris, France}
\affiliation{Sorbonne Universités, UPMC Univ. Paris 06, Universit\'e Paris Diderot, Ecole normale sup\'erieure, CNRS, Laboratoire de Physique Statistique (LPS ENS), 75005 Paris, France}
\affiliation{Universit\'e Paris Diderot, Sorbonne Paris Cit\'e, Laboratoire de Physique Statistique de l'ENS (LPS ENS), Ecole normale sup\'erieure, UPMC Univ. Paris 06, CNRS, 75005 Paris, France}

\author{Varsha Dani}
\affiliation{University of New Mexico.}

\author{Guilhem Semerjian}
\affiliation{Laboratoire de physique th\'eorique, D\'epartement de physique de l'ENS, Ecole normale sup\'erieure, UPMC Univ. Paris 06, CNRS, PSL Research University, 75005 Paris, France}
\affiliation{Sorbonne Universités, UPMC Univ. Paris 06, Ecole normale sup\'erieure, CNRS, Laboratoire de Physique Th\'eorique (LPT ENS), 75005 Paris, France}

\author{Lenka Zdeborov\'a}
\affiliation{Institut de Physique Th\'{e}orique, CNRS, CEA, Universit\'{e} Paris-Saclay, F-91191, Gif-sur-Yvette, France.}

\date{\today}

\begin{abstract}%
     We study in this paper the structure of solutions in the random 
     hypergraph coloring problem and the phase transitions they undergo
     when the density of constraints is varied. Hypergraph coloring is a
    constraint satisfaction problem where each constraint includes $K$
    variables that must be assigned one out of $q$ colors in such a
    way that there are no monochromatic constraints, i.e. there are at
    least two distinct colors in the set of variables belonging to every
    constraint. This problem generalizes naturally coloring of random
    graphs ($K=2$) and bicoloring of random hypergraphs ($q=2$), 
    both of which were extensively studied in past works. 
    The study of random hypergraph
    coloring gives us access to a case where both the size $q$ of the 
    domain of the variables and the arity $K$ of the constraints
    can be varied at will. Our work provides explicit values and predictions 
    for a number of phase transitions
    that were discovered in other constraint satisfaction problems but
    never evaluated before in hypergraph coloring.  
    Among other cases we revisit the hypergraph bicoloring problem ($q=2$)
    where we find that for $K=3$ and $K=4$ the colorability threshold
    is not given by the one-step-replica-symmetry-breaking analysis as
    the latter is unstable towards more levels of replica symmetry
    breaking. We also unveil and discuss the coexistence of two different
    1RSB solutions in the case of $q=2$, $K \ge 4$. Finally we present asymptotic
    expansions for the density of constraints at which various phase 
    transitions occur, in the limit where $q$ and/or $K$ diverge.

\end{abstract}

\pacs{Classification, Keywords}

\maketitle

\section{Introduction}
\label{sec:intro}
Constraint satisfaction problems (CSPs) where values are assigned to a set of variables in such a way to satisfy a given set of constraints arise in many areas of science. In particular they stand at the core of the theory of computational complexity~\cite{cook1971complexity} and analysis of algorithms. The study of random constraint satisfaction problems became popular in the endeavor to analyze the computational complexity of typical instances, instead of the worst possible case~\cite{cheeseman1991really,mitchell1992hard}. In random CSPs the set of constraints is chosen from some simply defined probability distribution, typical instances being those sampled with high probability from this distribution. Random CSPs are formally closely related to models considered in statistical physics of disordered systems and spin glasses~\cite{MezardParisi87b,MezardMontanari07}, and the application of statistical mechanics methods to random CSPs turned out to be particularly fruitful.

The most widely studied random CSP is satisfiability of random Boolean formulas ($K$-SAT). A series of works coming from statistical physics brought an explicit prediction for the so-called satisfiability threshold -- a density of constraints beyond which the formula is with high probability unsatisfiable~\cite{MezardParisi02,Mertens2005}. These studies also unveiled a peculiar structure of the satisfiable assignments in the satisfiable phase and additional phase transitions, besides the satisfiability one, at which the qualitative structure of the set of solutions changes drastically~\cite{MonassonZecchina99b,BiroliMonasson00,MezardParisi02,Krzakala2007}. This understanding inspired the development of a new class of message-passing-based satisfiability solvers that outperform existing ones for randomly generated formulas~\cite{MezardParisi02}. While many of these results were originally obtained using non-rigorous methodology (namely the cavity and the replica method), in recent years a large part of these results and the associated picture was established rigorously, see e.g.~\cite{mezard2005clustering,AchlioptasMoore06,AchlioptasRicci06,AchlioptasCoja-Oghlan08,ding2014proof}, leading to a new methodology in probability theory able to deal with probability distributions associated to randomly generated constraints on a large set of Boolean variables.

Studies of random CSPs were of course not restricted to random $K$-SAT, the statistical mechanics methodology being very versatile and easily adapted to other type of constraints and of variables. For instance random graph coloring is an eminent example in which the variables can take more than two values~\cite{mulet2002coloring,Krzakala2004,Zdeborova2007}. On the other hand in graph coloring every constraint corresponds to an edge and hence contains only two variables. Some of these results have been rigorously established, see e.g. in \cite{Sly08,molloy_col_freezing,coja2013chasing,bapst2016condensation,Sly16}, although the larger size of variables domain does complicate the situation. Some of the techniques are simpler for a version of the random graph coloring problem where every edge forbids a random set of neighboring colors~\cite{dani2012tight}. This ``permutated'' version of the coloring problem is conjectured to be equivalent to the usual one in terms of the structure of solutions and the phase transitions. However, proving this rigorously remains an open problem.

Another popular random CSP with Boolean variables and $K$-ary constraints is Not-All-Equal satisfiability (NAE-SAT) where each constraint contains $K$ random literals (i.e. variables randomly negated or not). This problem possesses a symmetry that is absent in random $K$-SAT and renders the problem simpler to analyze. A conjecture that follows from the statistical physics analysis is that the structure of solutions and associated phase transitions in random NAE-SAT are the same for bicoloring of random hypergraphs, where each constraint contains $K$ variables and states that they should not all take the same value. Hypergraphs generalize the concept of a graph in the sense that hyperedges connect to an arbitrary number of variables ($K$ in our case) whereas in graph an edge connects two variables. Statistical physics studies of the hypergraph bicoloring can be found in~\cite{CastellaniNapolano03,Dallasta2008,BrDaSeZd16}. Again, large part of the resulting picture was established rigorously~\cite{AchlioptasMoore06,AchlioptasCoja-Oghlan08,CojaZdeb12, CoPa12,DiSlSu13_naeksat,molloy_csp_freezing,BaCoRa14}.

Hypergraph coloring is another very natural CSP that generalized both graph coloring and hypergraph bi-coloring. Coloring of hypergraphs as defined in \cite{Krivelevich1998} consists of requiring that in the set of $K$ nodes belonging to a given constraint each of the $q$ colors appears at most $\gamma$ times, with $1 \le \gamma \le K-1$. The authors of~\cite{Dyer2015,Ayre2015} then analyzed coloring of random hypergraphs for $\gamma=K-1$, i.e. the constraints only forbid configurations where all the nodes that belong to one constraint have the same color. This is also the version of the problem we consider in the present paper, in which the statistical mechanics methodology is applied for the first time to the random hypergraph coloring with arbitrary values of $K$ and $q$, yielding results on the structure of solutions and phase transitions comparable to the state-of-the-art for random graph coloring and random $K$-SAT. 

Besides the quantitative computation of the thresholds for various phase transitions in this problem, we shall present a number of qualitatively new results. While the bicoloring ($q=2$) problem was studied previously in a number of papers, we found indeed some behaviors and results that were missed. Our first main result concerns the fact that on Erd\H{o}s-R\'{e}nyi random $K$-uniform hypergraphs with $K=3$ and $K=4$ the colorability threshold is not given by the one-step-replica-symmetry breaking (1RSB) analysis. An explicit prediction for the colorability threshold in those cases thus remains an open problem. This contrasts with the conjectures existing for random graph coloring, and random satisfiability \cite{Krzakala2004,MontanariParisi04}. The second main result is the coexistence of two distinct 1RSB solutions for $q=2$ and $K=4$ in the satisfiable region. Finally, another important outcome of this paper consists in asymptotic expansions for the phase transition thresholds in a model where both the parameters $K$ and $q$ can diverge, as far as we know no such case was discussed previously in the statistical mechanics literature.

The rest of the paper is organized as follows. In Sec.~\ref{sec:Ph} we define precisely the model under study, phrase it as a statistical mechanics problem and discuss qualitatively the various phase transitions it undergoes. In Sec.~\ref{sec:CavityEq} we derive the equations that describe the problem in the formalism of the cavity method (including the phenomenon of replica symmetry breaking). Our main results are presented in Sec.~\ref{sec:results}, where we discuss the numerical resolution of these equations, as well as their asymptotic expansions for large $K$ and/or $q$. We present our conclusions and perspectives for future work in Section~\ref{sec:conclu}, while a series of Appendices contain some more technical details.

\section{Statistical physics formulation of hypergraph $q$-Col}
\label{sec:Ph}
\subsection{Model}
\label{subsec:model}
\subsubsection{Constraint satisfaction problems as interacting spin models}
In a constraint satisfaction problem (CSP) $N$ variables $\{x_i\}_{i=1}^N$, each taking values in a finite set $\mathcal{Q}$, are required to satisfy a set of $M$ constraints. The $\mu$-th constraint (also called clause, and indexed with $\mu \in \{1,\dots,M\}$) involves a subset $\partial \mu \subset \{1,\dots,N\}$ of the variables, and is defined by a function $\Delta_\mu (\mathbf{x}_{\partial\mu})$ equal to 1 if the constraint is satisfied, and 0 otherwise (here and in the following we denote $\mathbf{x}_S=\{x_i\}_{i \in S}$ the configuration of a subset $S$ of the variables). An assignment $\mathbf{x}=\br{x_1, ..,x_N}\in\mathcal{Q}^N$ of the variables is called a solution if it simultaneously satisfies all the constraints. A CSP is said to be satisfiable if it admits at least one solution.

Such a CSP can be conveniently represented by a factor graph of $N$ vertices and $M$ factors, drawn respectively as circles and squares in Fig.~\ref{fig:BP}. Each vertex $i$ carries the variable $x_i$, each factor $\mu$ is associated to the clause $\Delta_\mu$, an edge being drawn between a factor $\mu$ and a variable $i$ if and only if $\Delta_\mu$ depends on $x_i$. This gives a graphical interpretation of $\partial \mu$ as the variables adjacent to $\mu$ in the factor graph, and we shall denote similarly $\partial i$ the set of contraints the variable $x_i$ is involved in.

A bridge towards statistical mechanics is built by interpreting the variables $x_i$ as spins taking a finite number of distinct values ($\mathcal{Q} = \{-1,1\}$ for Ising spins, $\mathcal{Q} = \{1,\dots,q\}$ for Potts ones) and defining an Hamiltonian (or energy, or cost function) on the set of spin configurations as
\begin{align}
\Ha(\mathbf{x}) = \sum_{\mu = 1}^M 
\left(1 - \Delta_\mu (\mathbf{x}_{\partial \mu})\right) \, ,
\end{align}
which counts the number of unsatisfied constraints. Each spin interacts with its nearest neighbors along the CSP factor graph: a clause $\mu$ involving $K$ spins models a $K$-wise interaction, so that the energy shift for breaking the constraint is equal to $1$. Solutions of the CSP corresponds to zero energy configurations, hence an instance of a CSP is satisfiable if and only if its groundstate (minimal) energy is equal to zero. The Gibbs Boltzmann probability density over the configuration space associated to this Hamiltonian is
\begin{align}
\label{eq:gibbsmeas}
p(\mathbf{x}) = \frac{e^{- \beta \Ha(\mathbf{x}) }}{\Z(\beta)} \, ,
\end{align}
where $\beta$ is a parameter known as the inverse temperature and the partition function $\Z(\beta)$ normalizes this probability law. In the zero temperature limit ($\beta \to \infty$) $p$ concentrates on the optimal configurations minimizing the number of unsatisfied constraints. In particular if the instance is satisfiable $p$ becomes in this limit the uniform measure over the solutions, and $\Z$ counts then the number of these satisfying assignments.

\subsubsection{Hypergraph $q$-Col}
We shall consider in this paper the hypergraph $q$-coloring problem: it consists in assigning one color, among $q$ available, to each vertex $i$ so that none of the hyperedges $\mu$ is monochromatic (an hyperedge is said to be monochromatic if all the vertices it involves carry the same color).

This definition fits naturally in the framework defined above, by choosing the alphabet $\mathcal{Q}=\{1, 2, .., q\}$ to encode the possible colors of the vertices, denoting $N$ and $M$ the number of vertices and hyperedges of the hypergraph, and associating to each hyperedge $\mu$ a constraint function ${\Delta_\mu (\mathbf{x}_{\partial \mu}) = 1-\delta\left(\mathbf{x}_{\partial \mu}\right) }$, with $\delta(x_1,\dots,x_K) = 1$ if all elements in $x_1,\dots,x_K$ are equal and $\delta(x_1,\dots,x_K) = 0$ otherwise, in such a way that $\Delta_\mu$ equals 1 if and only if the $\mu$-th hyperedge is non-monochromatic. The Hamiltonian thus becomes
\begin{align}
\Ha(\mathbf{x}) = \sum_{\mu = 1}^M  \delta (\mathbf{x}_\mu) \, ,
\end{align}
a generalization of the usual Potts antiferromagnet from graphs ($K=2$) to hypergraph ($K \ge 3$) interactions.

\subsubsection{Random ensembles for typical case analysis}

As explained in the introduction random ensembles of instances provide useful benchmarks to assess the typical difficulty of a CSP; in the case of hypergraph $q$-coloring an instance is fully specified by an hypergraph, we shall thus consider ensembles of random hypergraphs, and in particular the following two distributions:

(i) the $K$-uniform $\ell$-regular random hypergraph, for which each hyperedge links $K$ vertices, and each vertex belongs to $\ell$ hyperedges, all the hypergraphs fulfilling these conditions being equiprobable in this ensemble. The numbers $N$ of vertices and $M$ of hyperedges must obviously obey the relationship $N \ell = M K$.

(ii) the $K$-uniform Erd\H{o}s - R\'enyi (ER) hypergraph, where $M$ hyperedges are chosen uniformly at random among all $\binom{N}{K}$ possible $K$-uplets of distinct variables. 

Explicit analytic results can be obtained for the typical properties of such random structures in the large size limit (called thermodynamic limit in physics jargon), in which both $N$ and $M$ go to infinity at a fixed ratio denoted $\alpha=M/N$. A crucial property shared by these two ensembles is the local convergence in this limit to an hypertree: the neighborhood within a fixed distance around an uniformly chosen vertex is, with a probability going to 1 in the thermodynamic limit, acyclic. This limit tree can be described in terms of a Galton-Watson branching process, in which the root has a number $d$ of neighboring hyperedges with probability $p_d$, each of the $d(K-1)$ vertices at distance 1 from the root giving themselves birth to a number of offsprings with a probability distribution denoted $r_d$, and so on and so forth. The excess degree distribution $r_d$ is the probability that one vertex in an uniformly chosen hyperedge has degree $d+1$, and is related to $p_d$ through
\begin{align}
\label{eq:rd}
  r_d = \frac{(d+1)p_{d+1}}{\underset{d'=1}{\overset{\infty}{\sum}} d' \, p_{d'}} \ .
\end{align} 

In the regular ensemble, case (i) above, all vertices have degree $\ell$, hence ${p_d = \delta_{d,\ell}}$ and ${r_d = \delta_{d,\ell-1}}$ with $\delta$ denoting here the Kronecker symbol. The ratio $\alpha$ of constraints by variables is given in terms of the parameters of this ensemble as $\alpha = \ell/K$.

In the Erd\H{o}s - R\'enyi case both $p_d$ and $r_d$ are found to be Poisson laws of average $\alpha K$. With a slight abuse of notation we shall denote $\ell = \alpha K$ the average degree in this case, there should be no possibility of confusion with the degree of the regular ensemble.

Some of our results will be obtained in the large degree limit (taken after the thermodynamic limit), in which a Poisson distributed random variable concentrates around its mean and hence the two ensembles become essentially equivalent, which will allow us to use the one in which the computations are the simplest.

\subsection{Phase transitions and thresholds}
\label{subsec:phaseTransitions}

The random ensembles of hypergraph $q$-coloring are parametrized by $q$, the number of colors, $K$ the arity of the constraints, and $\alpha$ the ratio of the number of constraints per variable (or equivalently $\ell = \alpha K$ the average degree of the variables). For a given choice of $q$ and $K$ the problem becomes more and more constrained as $\ell$ increases, and one can naturally expect the existence of a critical value $\ell_{\rm col}$ below which typical random hypergraphs are $q$-colorable, while with high probability they are un-colorable for $\ell > \ell_{\rm col}$. This satisfiability phase transition (or threshold phenomenon) is actually accompanied by a series of phase transitions affecting the structure of the space of solutions in the satisfiable phase $\ell < \ell_{\rm col}$. These structural phase transitions have been discovered by the applications of statistical mechanics methods, they are essentially the same as those discussed in the mean field theory of structural glasses~\cite{BeBi11}, and are largely independent of the details in the definition of the random CSP~\cite{Krzakala2007,MezardMontanari07}. In the next section we first explain these phase transitions in a qualitative way, before moving gradually to a more quantitative and technical description.

\subsubsection{Qualitative picture of the phase diagram}
\label{sec:qualitative_picture}

At low density of constraints all solutions are found in one ``connected'' portion of configuration space, meaning that it is possible to go from any solution to any other solution by ``flipping'' only one (or a number sub-linear in $N$) variable at each step without breaking any constraint along the path. Equivalently, the set of solutions is an ergodic subspace and Monte Carlo Markov chains can achieve efficient sampling of solutions. 

As the density of constraints is increased a first remarkable transition is encountered: at the \emph{clustering} transition the set of solutions decomposes in an exponential (in $N$) number of ``disconnected'' smaller sets,  called \emph{clusters}. Following the physics formulation of CSPs presented in the previous section, there are either energetic or entropic barriers between different clusters while there exist zero energy paths between solutions in the same cluster. The precise definition of a cluster may vary among authors of previous works, we shall clarify our operational definition of this threshold $\ell_{\rm clust}$ in Sec. \ref{sec:CavityEq} while detailing our computations.

Further increasing $\ell$ beyond $\ell_{\rm clust}$, a \emph{condensation} transition occurs at $\ell_{\rm cond}$: it separates a regime in which an exponential number of clusters contain most of the solutions, to a situation for $\ell > \ell_{\rm cond}$ in which most of the solutions are found in an only sub-exponential number of clusters. 

Eventually, the disappearance of the last surviving clusters of solutions marks the \emph{colorability} (or \emph{satisfiability} for a generic CSP) threshold, at $\ell_{\rm col}$. 

The partition of the set of solutions into clusters allows to define the notion of the \emph{frozen} variables of a solution: these are the variables which take the same color in all the solutions of the corresponding cluster. One can then further refine the description of the clustered phase and introduce two new phase transitions: the \emph{rigidity} transition, denoted $\ell_{\rm r}$, above which the typical solutions contain an extensive number of frozen variables, and the \emph{freezing} transition $\ell_{\rm f}$, above which all solutions have this property (this latter threshold is much more difficult to compute as it requires a control of all solutions, including atypical ones, and has been estimated analytically only for the bicoloring of hypergraphs~\cite{BrDaSeZd16}).

As various notations have been used across different works, we recap the definition of the commonly encountered thresholds in the study of random CSPs along with our notations in Appendix \ref{app:thresholds} . 

By definition of the different thresholds the inequalities $\ell_{\rm clust}\leq \ell_{\rm cond} \leq \ell_{\rm col}$ and $\ell_{\rm clust}\leq \ell_{\rm r} \leq \ell_{\rm col}$ must hold. A common phenomenon observed in many CSPs, such as $q$-coloring~\cite{Zdeborova2007}, bicoloring of $K$-regular graphs~\cite{Dallasta2008,BrDaSeZd16}, or $K$-SAT~\cite{MoRiSe08}, is the saturation of some of these bounds when the parameters $q$ or $K$ are small (for instance $\ell_{\rm clust} = \ell_{\rm cond}$ in the 3-coloring of random graphs~\cite{Zdeborova2007}). On the contrary when $q$ or $K$ are sufficiently large the generic order for the thresholds becomes $\ell_{\rm clust} <  \ell_{\rm r} < \ell_{\rm cond} < \ell_{\rm col}$. Additionally, asymptotic expansions performed in the limit of diverging $q$ or $K$~\cite{Mertens2005,Zdeborova2007} revealed the existence of two main scales for these various thresholds, namely $\ell_{\rm clust} \sim \ell_{\rm r} \ll \ell_{\rm cond} \sim \ell_{\rm col}$.

Note that the asymptotic expansions of the thresholds predicted by the statistical mechanics methods have been performed up to now in models in which only one parameter could be taken to infinity ($K$ was fixed to 2 for the $q$-coloring of random graphs, and $q$ was fixed to 2 for the bicoloring of random hypergraphs). In the presently studied model one can perform asymptotic expansions in which $q$ and/or $K$ can be taken to infinity. We will come back in Sec.~\ref{sec:asymptotic} on these expansions and on the interplay between these two possible limits.

In the following section we review how the cavity approach reveals the location of the different transitions.

\subsubsection{Overview of the cavity approach and key observables}
\label{sec:overview_cavity}
A central objective of the statistical mechanics approach to random CSP is the computation of the quenched free-entropy density
\begin{align}
\label{eq:phiquenched}
\phi(\beta) = \lim_{N \to \infty} \frac{1}{N}\mathbb{E}[ \ln \Z (\beta) ] \ ,
\end{align}
where $\Z$ is the partition function normalizing the probability law \eqref{eq:gibbsmeas}, and the average is over the random hypergraph ensemble. Indeed the free-entropy $\ln \Z$ is expected to concentrate around its mean (a property called self-averaging in physics), which is thus representative of the properties of the typical hypergraphs of the ensemble. The behavior of $\phi$ in the zero-temperature limit $\beta \to \infty$ depends on whether the typical graphs are colorable or not. If they are the limit of $\phi$, to be denoted $s$, is the exponential rate of growth of the number of solutions, called entropy. If they are not $\phi$ diverges with $\beta$, with a proportionality constant that gives the minimal number of unsatisfied constraints. For simplicity in the rest of this section we assume that we are in the satisfiable phase and we keep the zero-temperature limit implicit, the reader can adapt this discussion to the positive temperature case without difficulties.

The belief propagation (BP) algorithm~\cite{KschischangFrey01} provides a tractable procedure (see Sec. \ref{sec:CavityEq}) to compute the Bethe approximation of the entropy density $s = \ln \Z / N$ for a given hypergraph. The replica symmetric (RS) cavity method is a way to average this prediction over the random hypergraph ensemble in the thermodynamic limit. BP yields an exact result for a tree factor graph. As we have seen above the random graph ensembles studied in this paper, with the sparsity assumption we made keeping $\alpha = M/N$ finite, are locally tree-like, their shortest loops are typically of length $O(\ln N)$. As a consequence the predictions of BP are asymptotically exact provided the longer loops do not induce long range correlations between far away variables. This is precisely the case in the unclustered phase $\ell < \ell_{\rm clust}$, but becomes wrong for larger average degrees.

The clustered structure of the configuration space causes indeed the birth of a form of correlation between variables (measured by a so called point-to-set correlation function), the connection between these two properties being reminiscent of the link between spatial and temporal mixing in the studies of Monte Carlo Markov Chain dynamics. It is thus necessary to correct the RS cavity method above the clustering threshold, by promoting it to the Replica Symmetry Breaking (RSB) cavity method. A crucial idea of this method is to exploit the partition of the solution space into disjoint clusters $\br{\C}$, assuming that the measure \eqref{eq:gibbsmeas} restricted to the solutions $\mathbf{x}$ of one cluster $\C$ enjoy the decorrelation property described above. This restricted measure can thus be treated within the RS formalism, and what remains to do is to describe the statistical properties of the cluster decomposition. More precisely, this revised formalism introduces a measure over the set of clusters $\br{\C}$,
\begin{align}
\label{eq:p1}
p_1 \p{\C} = \frac{\Z(\C)^m}{\Z_1(m)} = \frac{e^{mNs(\C)}}{\Z_1(m)} \; ,
\end{align} 
where $\Z(\C)$ and $s(\C)=\ln \Z(\C)/N$ are respectively the number of solutions in $\C$ and the \emph{internal} entropy of $\C$, and where $m$ is a parameter playing a role similar to the temperature called the \emph{Parisi parameter}.
The 1RSB normalizing function $\Z_1$ is thus given by
\begin{align}
\label{eq:defZ1}
\Z_1(m) = \sum_{\br{\C}} e^{mNs(\C)} = \integ{s} e^{N(ms + \Sigma (s))} \,
\end{align}
where we introduced the \emph{complexity} $\Sigma (s)$, defined as the logarithm of the average number of clusters of internal entropy $s$ (divided by $N$). The saddle point evaluation of this integral yields a thermodynamic potential that we shall call \emph{replicated entropy},
\begin{align}
\label{eq:defPhis}
\Phi(m) =\lim_{N \to \infty} \frac{1}{N} \ln \Z_1(m) =  m \, s(m) + \Sigma (m) \, ,
\end{align}
where we defined
\begin{align}
\label{eq:defsm}
s(m) = \underset{s}{\operatorname{argmax}}\s{m\, s+\Sigma(s)},
\end{align}
and $\Sigma(m) = \Sigma(s(m))$.

Exploiting the decomposition of the solution set into clusters we can express the total entropy as
\begin{eqnarray}
\stot &=& \lim_{N \to \infty} \frac{1}{N} \ln \integ{s} e^{N(s + \Sigma (s))} \nonumber \\ 
&=& \sup_{s / \Sigma(s)\geq 0} \s{s + \Sigma(s)} \, ,
\label{eq:defstot}
\end{eqnarray}
where the second equality results again from a saddle-point evaluation, the condition $\Sigma(s) \geq 0$ ensuring that the corresponding clusters exist in the thermodynamic limit. The total entropy thus results from a competition between most numerous clusters (that have a large complexity) and bigger clusters (that have a large internal entropy). Within the exponential approximation the total entropy is dominated by the contribution of the clusters of internal entropy hereafter called $\sstar$ achieving the supremum of Eq. \eqref{eq:defstot}, we thus call a solution typical if it belongs to a cluster of internal entropy $\sstar$.

If $\Sigma(\sstar) > 0$, there is an exponential number of clusters contributing to the total entropy, we are thus in the \emph{clustered phase} as defined in the previous section. On the contrary if $\Sigma(\sstar) = 0$,  either it corresponds to connectivities smaller than $\ell_{\rm clust}$ where there is only one giant cluster, or more interestingly, to the \emph{condensed phase}, where a sub-exponential number of clusters contains the vast majority of solutions.
 
The computation of the complexity at $m=1$ gives access to  $\ell_{\rm clust}$ and $\ell_{\rm cond}$: in \eqref{eq:defsm} the maximization is not constrained by the condition $\Sigma(s) \ge 0$, hence the total entropy $\stot$ is equal to $\Phi(m=1)$ if and only if $\Sigma(m=1)>0$, while $\Sigma(m=1) < 0$ is the signature of the condensation phenomenon. In the clustered un-condensed regime the prediction $\Phi(m=1)$ actually coincides with the RS prediction: the superposition of the exponentially large number of clusters create relatively weak correlations, that can be only detected by a point-to-set correlation function and not correlation between a finite number of variables, see~\cite{CoKrPeZd16,CoEfJaKaKa17} for recent rigorous proofs of the exactness of the RS predictions up to the condensation phase transition.

Another equivalent and compact way of summarizing the thermodynamics prescription is the formula:
\begin{align}
\label{eq:min_Phi}
s_{\rm tot} = \inf_{m \in [0,1]} \frac{\Phi(m)}{m} \ ,
\end{align}
which is valid irrespectively of the type of phase: in an unclustured situation, with a single cluster of entropy $s_{\rm tot}$, equation \eqref{eq:defZ1} shows that $\Phi(m)=m s_{\rm tot}$, hence the parameter $m$ is irrelevant in the right hand side of \eqref{eq:min_Phi}. In presence of clusters the minimizer in \eqref{eq:min_Phi}, that we shall call $m_{\rm s}$ for the \emph{static} value of the Parisi parameter, is either $m_{\rm s}=1$ if an exponential number of clusters contribute to the total entropy, or some value $m_{\rm s}<1$ if the measure condensates on a sub-exponential number of clusters. Indeed, one can see that $\frac{\rm d}{{\rm d}m} (\Phi(m)/m) = -\Sigma(m)/m^2$, hence in the condensed phase the minimization selects the largest existing clusters, with $\Sigma(m_{\rm s})=0$, as in \eqref{eq:defstot}. 

Finally the colorability threshold can be obtained from the computation of $\Phi(m=0)=\Sigma(m=0) = \max_s \Sigma(s)$: this gives the exponential rate of growth for the total number of clusters of solutions, irrespectively of their sizes. The colorability transition occurs when this number vanishes, marking the disappearance of the last clusters of solutions.

\section{Cavity equations and thresholds characterizations}
\label{sec:CavityEq}
In this section we derive the cavity equations of hypergraph $q$-coloring at finite temperature $\beta$. The zero temperature equations follow straightforwardly in the limit $e^{-\beta}\rightarrow 0$. These equations cover the special cases of graph $q$-coloring (when $K=2$) \cite{Krzakala2004,Zdeborova2007}, and hypergraph bi-coloring or NAESAT (when $q=2$) \cite{CastellaniNapolano03,Dallasta2008,BrDaSeZd16}.

\subsection{Replica symmetric formalism}

\subsubsection{Cavity equations and Bethe entropy}
\label{subsec:RScavity}

\begin{figure}
\begin{center}
      \includegraphics[width=0.25\textwidth]{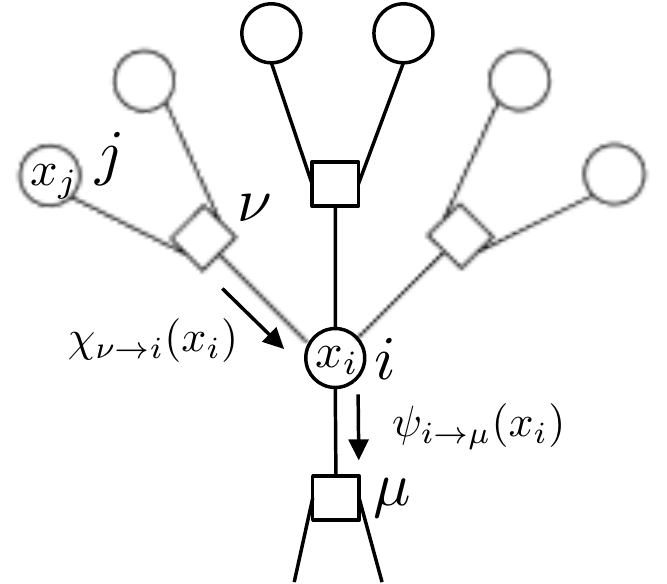}
     \caption{Sketch of a local architecture of hypergraph illustrating notations for the derivations of the cavity equation. Vertices are represented with circles, and hyperedges with squares.\label{fig:BP}}
\end{center}
\end{figure}

\paragraph{Messages}

Consider first the problem of computing the partition function $\Z(\beta)$ and characterizing the marginals of the probability law $p(\mathbf{x})$ defined in \eqref{eq:gibbsmeas} when the factor graph encoding the dependencies between variables is a tree. Then the problem can be solved exactly and very easily by exploiting the recursive nature of trees, with a procedure called dynamic programming or Belief Propagation~\cite{KschischangFrey01}. 

We denote by $\X \nti \p{x_i}$ the marginal probability of the variable $x_i$ in the law encoded by the amputated factor graph in which all edges around $i$ have been cut except the one connecting it to $\nu$, and similarly $\Y\itm \p{x_i}$ for the amputated factor graph where only the edge between $i$ and $\mu$ has been cut (see the illustration in Fig.~\ref{fig:BP}). If the factor graph is a tree these ``messages'' sent between neighboring variable and factor nodes are easily seen to obey the following exact equations:
\begin{align}
\label{eq:genedgmsg}
\X \nti \p{x_i} & = \frac{1}{\hZ\ntiu_0}\sum_{\mathbf{x}_{\partial \nu \setminus i}} e^{-\beta(1 - \Delta_\nu(\mathbf{x}_{\partial \nu }))} \prod\nbe{j}{\nu}{i} \Y\jtn \p{x_j} \, , 
\end{align}
and
\begin{align}
\label{eq:ndmsg}
\Y\itm \p{x_i} = \frac{1}{\hZ\itmu_0}   \prod\nbe{\nu}{i}{\mu} \X\nti(x_i) \, ,
\end{align}
where we recall that $\partial i$ and $\partial \mu$ denote respectively the set of factor nodes around variable $i$ and the set of variable nodes around factor $\mu$, and $\hZ\ntiu_0$ and $\hZ\itmu_0$ are normalization factors ensuring that both $\X \nti $ and $\Y\itm$ sum up to one, i.e.
\begin{align}
\sum_{x = 1}^q \X \nti \p{x} = \sum_{x = 1}^q \Y\itm \p{x} =1 .
\end{align} 
Equation \eqref{eq:genedgmsg} has been written for a generic CSP; specializing it to the hypergraph $q$-coloring case yields
\begin{align}
\label{eq:edgmsg}
\X \nti \p{x_i}  =  \frac{1}{\hZ\ntiu_0} \p{1 + (e^{-\beta}-1) \prod\nbe{j}{\nu}{i} \Y\jtn \p{x_i}}  \; ,
\end{align}
as given $x_i$ the only penalized configuration is the one in which all neighbors $j$ of clause $\nu$ take this same color $x_j = x_i$.

From \eqref{eq:ndmsg} and \eqref{eq:edgmsg} we deduce a closed recursion on messages from nodes to hyperedges
\begin{align}
&\Y\itm \p{x_i} = \frac{1}{\Z\itmu_0} {\displaystyle\prod\nbe{\nu}{i}{\mu} \p{1 + (e^{-\beta}-1) \prod\nbe{j}{\nu}{i} \Y\jtn \p{x_i} } }
     \;, \nonumber \\
&\Z\itmu_0 = {\displaystyle\sum_{x=1}^q \prod\nbe{\nu}{i}{\mu} \p{1 + (e^{-\beta}-1) \prod\nbe{j}{\nu}{i} \Y\jtn \p{x} }} \;,
\label{eq:defZ0}
\end{align} 
and define the shorthand notation
\begin{align}
\label{eq:closedmsg}
\Y\itm    \equiv  \FRS (\br{\Y\jtn} _{\nu \in \partial i \setminus \mu ,j \in \partial \nu \setminus i})
    \equiv \FRS  (\br{\Y\jtn}) \; ,
\end{align} 
for this relation.

The marginal probability of variable $x_i$ in the full factor graph can then be obtained from the messages as
\begin{align}
p_i \p{x_i}   & = \frac{1}{\Z^i_0} \prod\nb{\nu}{i} \X\nti(x_i) \, ,
\end{align} 
where $\Z^i_0 $ is the marginal partition function
\begin{align}
\Z^i_0    & = \sum_{x=1}^q \prod\nb{\nu}{i} \X\nti(x) \; .
\end{align}

If the factor graph is a tree there exists a single solution to the consistency equations between messages, and the prediction thus obtained for the marginals of $p(\mathbf{x})$ is exact. One can however look for a fixed point of Eq. \eqref{eq:closedmsg} on an arbitrary hypergraph even if it has loops: this is precisely the BP algorithm, that we expect to converge for typical random hypergraphs of small enough average degree.

The replica symmetric cavity method provides a way to deal with such random (hyper)graphs ensembles. Consider the thought experiment in which one generates at random a large hypergraph with a degree distribution $p_d$, finds the fixed point of BP on this particular hypergraph (assuming it exists and is unique), and choose uniformly at random a directed edge $i \to \mu$ in this hypergraph. The message $\Y\itm$ thus obtained is random, and we denote $\Pm^{\rm RS}(\Y)$ its probability distribution. For consistency reasons this distribution has to obey the following equation
\begin{align}
\label{eq:pdfmsg}
\Pm^{\rm RS}(\Y) = \sum_{d=0}^{ \infty} r_d 
\int \prod_{\nu = 1}^d \prod_{j=1}^{K-1} &
 \mathrm{d} \Y\jtn \Pm^{\rm RS}(\Y\jtn) \notag \\
& \kr{\Y}{\FRS(\br{\Y\jtn})} \;,
\end{align}
where $\delta[]$ is the Dirac distribution, and $r_d$ the excess degree distribution associated to $p_d$ according to \eqref{eq:rd}.

\paragraph{Bethe free energy}

On a tree factor graph the messages solution of \eqref{eq:closedmsg} can be used to compute exactly the associated partition function. One finds indeed that
\begin{align}
\label{eq:deffB}
\phiB=
\frac{1}{N} \ln \Z =
\frac{1}{N} \sum_{i=1}^N \ln \Z^{i+\partial i}_0 - \frac{(K-1)}{N} \sum_{\mu=1}^M \ln\Z^\mu_0 \, , 
\end{align}
where in the general CSP formalism
\begin{align}
\label{eq:gendefZo}
\Z_0^\mu    & = \sum_{\mathbf{x}_{\partial \mu}} e^{-\beta(1-\Delta_\mu(\mathbf{x}_{\partial \mu }))} \prod\nb{i}{\mu} \Y\itm \p{x_j}  \, , \\
\Z_0^{i+\partial i}   & = \sum_{x_i=1}^q \prod\nb{\mu}{i}  
        \sum_{\mathbf{x}_{\partial \mu \setminus i}} e^{-\beta(1-\Delta_\mu(\mathbf{x}_{\partial \mu }))} \prod\nbe{j}{\mu}{i} \Y\jtm \p{x_j} \, ,\label{eq:gendefZidi}
\end{align}
which yields for the hypergraph $q$-coloring interactions
\begin{align}
\Z_0^\mu    &  = 1 + (e^{-\beta}-1) \sum_{x =1}^q \prod\nb{i}{\mu}  \Y\itm \p{x} \, , \label{eq:edgRSprt} \\
\Z_0^{i+\partial i}   & = \sum_{x=1}^q \prod\nb{\mu}{i}  \p{1 + (e^{-\beta}-1) \prod\nbe{j}{\mu}{i} \Y\jtm \p{x}} \,.
\end{align}
If the factor graph is not a tree, but if BP converges, one can use the same expressions with the messages solution of the BP equations: this yields the Bethe approximation for the free entropy density.

The prediction of the cavity method at the RS level for the quenched free entropy density $\phi(\beta)$ defined in \eqref{eq:phiquenched} is then obtained by averaging this Bethe approximation with respect to the message distribution $\Pm^{\rm RS}(\Y)$ and the degree distribution $p_d$, yielding
\begin{widetext}
\begin{align}
\phi^{\rm RS}(\beta) =  \sum_{d=0}^{\infty}  p_d \int \prod_{\mu=1}^d\prod_{j=1}^{K-1} \mathrm{d} \Y\jtm \Pm^{\rm RS}(\Y\jtm) 
\ln \Z_0^{i+\partial i}(\{\Y\jtm\})
- \frac{\ell(K-1)}{K} \int \prod_{i=1}^K \mathrm{d} \Y_i \Pm^{\rm RS}(\Y_i) 
\ln \Z_0^\mu(\psi_1,\dots,\psi_K)
\label{eq:fBqdvlp}\; ,
\end{align}
\end{widetext}
with $\ell$ the average of $p_d$.

\subsubsection{Replica symmetric solution at zero temperature}

The replica symmetric cavity equation \eqref{eq:closedmsg} admits as a trivial fixed point, independently of the precise hypergraph, the uniform distribution,
\begin{align}
\forall i, \mu, x_i \quad \Y\itm(x_i )= \X\mti(x_i) = \overline{\psi} (x_i) = \frac{1}{q} \; .
\end{align}
It is actually the relevant one: as the interactions are antiferromagnetic the symmetry between colors cannot be broken in a pattern compatible with the existence of loops of all lengths in the hypergraph. 

The RS prediction for the quenched entropy at zero temperature is thus found by inserting the solution $\Pm^{\rm RS}(\Y) = \delta[\Y - \overline{\psi}]$ and setting $e^{-\beta}=0$ in \eqref{eq:fBqdvlp}, yielding
\begin{align}
\label{eq:defRSent}
s^{\rm{RS}}= \ln q + \frac{\ell}{K} \ln \p{1 - \frac{1}{q^{K-1}}} \, .
\end{align}
This expression is actually equal to the annealed entropy $\lim_{N \to \infty} \frac{1}{N}\ln \mathbb{E}[ \Z (\beta=\infty) ] $ counting the average number of solutions, and is hence an upper bound on the true quenched entropy. As a consequence the average degree above which the RS entropy is negative is an upper bound on the coloring threshold 
\begin{align}
\label{eq:lRS}
\ell_{\rm{RS}} = - K \ln q /\ln \p{1-1/q^{K-1}} > \ell_{\rm{col}} \; .
\end{align}
This corresponds to the first moment argument extensively used in computer science to obtain such upperbounds on the satisfiability transition, see for instance~\cite{Krivelevich1998,Dyer2015,Ayre2015} for its use in the context of hypergraph $q$-coloring.

\subsection{One step replica symmetry broken formalism}
\label{subsec:1RSBcavity}
As already discussed in Sec. \ref{subsec:phaseTransitions} the hypothesis underlying the RS formalism are violated for sufficiently large average degree, as the clustering phenomenon induce long-range correlations between variables that are neglected at the RS level. We shall now describe a more elaborate version of the cavity method, called 1RSB for ``one step of replica symmetry breaking'', that is able to tackle this clustering phenomenon, and then explain how to quantitatively determine the clustering, condensation and colorability thresholds.

\subsubsection{Cavity equations}

The crucial hypothesis of the 1RSB cavity method is that the decomposition of the Gibbs measure \eqref{eq:gibbsmeas} in the clusters partitioning the configuration space is a ``pure state'' decomposition in the mathematical physics sense, namely that the restriction of the Gibbs measure to one cluster $\C$ enjoy the decorrelation properties used at the RS level. Hence each cluster $\C$ is described by a set of BP messages $\{\Y\itm^\C \}$, solution of the BP fixed-point equations:
\begin{align} 
\Y\itm^\C=  \FRS  (\br{\Y\jtn^\C}) \; .
\end{align} 
Our goal is now to compute the 1RSB partition function \eqref{eq:defZ1} that sums the contribution of the different clusters to describe the thermodynamics of the space of solutions. As we assume that each cluster $\C$ is well described by a BP fixed point, we can use the Bethe approximation to compute the partition function restricted to $\C$ and write
\begin{align}
\label{eq:recalldefZ1}
\Z_1(m) = \sum_{\br{\C}} e^{mN\phiB(\C)} \; ,
\end{align}
where the internal free entropy density $\phiB(\C)$ is $\phiB(\br{\Y\itm^\C})$ defined in Eq. \eqref{eq:deffB}. Using this correspondence between clusters and BP fixed points, $\Z_1$ can be rewritten as an integral over BP messages:
\begin{widetext}
\begin{align}
\label{eq:rewriteZ1}
\Z_1(m)  = \int \prod_{(i\mu)}   {\rm d} \Y\itm 
\left(\prod_{(i\mu)} \kr{\Y\itm}{\FRS\p{\br{\Y\jtn}}} \right)
e^{mN\phiB(\br{\Y\itm})} \, ,
\end{align}
\end{widetext}
where $(i\mu)$ spans all the edges of the factor graph, and we used the shorthand notation $\FRS$ defined in \eqref{eq:closedmsg}.
The Dirac distributions in Eq. \eqref{eq:rewriteZ1} make the integral equivalent to the former sum over clusters of Eq. \eqref{eq:recalldefZ1}, by selecting only the messages that are BP fixed points.
Using the expression of $\phiB$ given in \eqref{eq:deffB}
we can interpret Eq. \eqref{eq:rewriteZ1} as the partition function of an auxiliary statistical physics problem~\cite{MezardMontanari07} where now the variables are the messages $\{\psi\itm\}$ instead of the spin variables $\{x_i\}$, and the interactions arise from the partition functions $\Z_0^{i+\partial i}$ and $\Z_0^\mu$ in $\phiB$. The factor graph encoding this auxiliary problem on the variables $\{\psi\itm\}$ is as locally tree-like as the original graph, one can thus treat it with the Belief Propagation approach explained above, with the messages in this auxiliary problem becoming probability distributions $P\itm \p{\Y\itm}$ over the variables $\{\psi\itm\}$. The 1RSB equations are nothing but the BP equations of this auxiliary problem, that read
\begin{widetext}
\begin{align}
\label{eq:nd1RSB}
P\itm\p{\Y\itm}  = \frac{1}{\Z\itmu_1} & \int
         \p{ \prod\nbe{\nu}{i}{\mu}\prod\nbe{j}{\nu}{i}  {\rm d} \Y\jtn P\jtn \p{\Y\jtn} }
         \delta \s{ \Y\itm - \FRS \p{ \br{\Y\jtn} }} 
         \p{\Z\itmu_0(\br{\Y\jtn} )}^m
         \, ,
\end{align}
\end{widetext}
where $\Z\itmu_1$ is a normalization that depends on the $\{P\jtn \}$ and on $m$; we shall write in short \eqref{eq:nd1RSB} as
\begin{align}
\label{eq:closedmsgmsg}
P\itm & \equiv \FRSB \p{\br{P\jtn}}.
\end{align}
The probability distributions $P\itm$ have to be interpreted as the probability to observe a BP message $\Y\itm^\C=\Y\itm$ when a cluster $\C$ is chosen randomly. As explained above we do not expect the symmetry between colors to be spontaneously broken, hence we impose that on average over the various clusters no color is privileged at any vertex, i.e.
\begin{align}
\int  \mathrm{d} \Y\itm P\itm\p{\Y\itm} \Y\itm = \overline{\psi}
\label{eq:P_sym}
\end{align}
where we had already defined $\overline{\psi}$ as the uniform distribution over $\mathcal{Q}$, with $\overline{\psi}(x)=1/q$.

\begin{widetext}
The Bethe free-entropy of this auxiliary model yields the 1RSB estimate of the replicated free-entropy defined in \eqref{eq:defPhis}, in terms of the auxiliary messages $\{P\itm \}$:
\begin{align}
\label{eq:RsB}
 \Phi^{\rm Bethe}(m)= \frac{1}{N}\sum_{i=1}^N \ln \Z_1^{i+\partial i} (m)
      - \frac{K-1}{N} \sum_{\mu=1}^M \ln \Z_1^{\mu} (m) \, ,
\end{align}
where
\begin{align}
\label{eq:defZ1idi}
\Z_1^{i+\partial i}(\br{P\jtn} ; m)  = \int \prod\nb{\mu}{i}
\prod\nbe{j}{\mu}{i} {\rm d} \Y\jtm 
P\jtm \p{\Y\jtm} \p{\Z^{i+\partial i}_0}^m
\end{align}
and
\begin{align}
\label{eq:defZ1mu}
\Z_1^{\mu} (\br{P\itm} ; m ) = \int \prod\nb{i}{\mu} 
{\rm d} \Y\itm 
P\itm \p{\Y\itm} \p{\Z^\mu_0}^m \; .
\end{align}
In order to deal with random graph ensembles we introduce a probability distributions of 1RSB messages $\Pm^{\rm{1RSB}}(P)$ that obey the analog of \eqref{eq:pdfmsg}, namely
\begin{align}
\label{eq:pdfmsgmsg}
\Pm^{\rm{1RSB}}(P) = \sum_{d=0}^{\infty} r_d  
\int \prod_{\nu = 1}^d \prod_{j=1}^{K-1} \mathrm{d} P\jtn \Pm^{\rm{1RSB}}(P\jtn) 
\kr{P}{\FRSB(\br{P\jtn})} \; .
\end{align}
From the solution of this equation one can compute the 1RSB estimate of the typical replicated free-entropy for a random hypergraph with degree distribution $p_d$. The formula is the analog of the one given in \eqref{eq:fBqdvlp} at the RS level, i.e.
\begin{eqnarray}
\Phi^{\rm 1RSB}(m) &=&  \sum_{d=0}^{\infty}  p_d \int \prod_{\mu=1}^d\prod_{j=1}^{K-1} \mathrm{d} P\jtm \Pm^{\rm 1RSB}(P\jtm) 
\ln \Z_1^{i+\partial i}(\{P\jtm\};m) \nonumber \\
&-& \frac{\ell(K-1)}{K} \int \prod_{i=1}^K \mathrm{d} P_i \Pm^{\rm 1RSB}(P_i) 
\ln \Z_1^\mu(P_1,\dots,P_K;m)
\label{eq:phi1RSB}\; .
\end{eqnarray}
\end{widetext}
The estimate of the complexity $\Sigma(m)$ is then deduced as
\begin{align}
\Sigma(m) = \Phi^{\rm 1RSB}(m) - m \frac{{\rm d}}{{\rm d}m} \Phi^{\rm 1RSB}(m) \, ,
\label{eq:SigmaLegendre}
\end{align}
following the discussion of Sec.~\ref{sec:overview_cavity}. To compute this derivative it is enough to take into account only the explicit dependency on $m$ of the functions $\Z_1$, as the potential \eqref{eq:phi1RSB} is stationary with respect to the variations of $\Pm^{\rm 1RSB}$ whenever the 1RSB cavity equation \eqref{eq:pdfmsgmsg} is fulfilled.

Note that the 1RSB formalism contains as a special case the RS one when the $P$'s in the support of $\Pm^{\rm{1RSB}}(P)$ are Dirac distributions; in the following we shall call this a trivial solution of the 1RSB equations. Such a solution expresses indeed the absence of clustering in the configuration space, as the randomness over the choice of the clusters becomes deterministic in this case.

\subsubsection{Hard fields and frozen variables}

We have introduced in Sec.~\ref{sec:qualitative_picture} the notion of frozen variables as the ones taking the same value in all the solutions of a cluster. In the cavity formalism a cluster is described in terms of BP messages, hence frozen variables are to be associated with ``hard fields'', i.e. messages that constrain one variable to a single color. More precisely, let us define $\psi^x$ the distribution over $\mathcal{Q}$ concentrated on the color $x$, i.e. $\psi^x(x')=\delta_{x,x'}$, and decompose the 1RSB distributions as
\begin{align}
P\itm\p{\Y\itm} &= \sum_{x=1}^q \eta\itm^x \delta[\Y\itm - \psi^x ] 
\label{eq:def_eta}\\ &+
 \eta\itm^0 \widetilde{P}\itm(\Y\itm) \, ,
\nonumber
\end{align}
where $\eta\itm^x$ is to be interpreted as the probability under $P\itm$ of an hard field imposing the color $x$, $\eta\itm^0=1-\sum_x \eta\itm^x$ is the complementary probability of ``soft fields'', and $\widetilde{P}\itm$ the distribution of the latter. When the symmetry between colors is respected we shall denote more simply $\eta\itm^x=\eta\itm$, then $\eta\itm^0=1-q\eta\itm$. Obviously this phenomenon of hard-fields can only occur at zero temperature; in the opposite case all configurations have positive probability, hence no marginals can be supported on a single color.

\subsubsection{Computation of the thresholds and of the free-entropy}
\label{sec:thresholds_computation}

Now that we have set up the cavity formalism at the 1RSB level we can come back on the definitions of the thresholds given in Sec.~\ref{subsec:phaseTransitions} and explain their practical computation:

\begin{itemize}

\item $\ell_{\rm clust}$ is the smallest average degree for which there exists a non-trivial solution of the 1RSB equations at $m=1$. An analysis of the local stability of the trivial RS solution yields an upperbound on this threshold (sometimes called the Kesten-Stigum bound), 
\begin{align}
\ell_{\rm clust} \le \ell^{\rm{RS}}_{\rm{stab}} = \frac{\p{q^{K-1}-1}^2}{K-1}
\label{eq:KSbound}
\end{align}
for Poisson degree distributions, as we shall prove in Appendix \ref{app:LinearStab_RS}. This bound is not tight in general because a bifurcation can cause the non-trivial solution to appear in a discontinuous way, far from the trivial solution, a phenomenon which cannot be detected by a local analysis.

\item $\ell_{\rm cond}$ is the smallest average degree for which there exists a non-trivial solution of the 1RSB equations at $m=1$ with $\Sigma(m=1)<0$.

\item $\ell_{\rm col}$ is the smallest average degree for which the complexity at $m=0$ is negative.

\item the prediction of the free-entropy that encompasses the RS and 1RSB formalism is 
\begin{align}
\inf_{m \in [0,1]} \frac{\Phi_{\rm 1RSB}(m)}{m} \ ,
\label{eq:min_Phi1RSB}
\end{align}
as follows from the 1RSB estimate of the potential $\Phi$ in the expression \eqref{eq:min_Phi} (for some models this prediction was proven rigorously to be an upperbound on $\phi$~\cite{FranzLeone03,PanchenkoTalagrand04}). At low degrees the 1RSB cavity equations only admits the trivial RS solution for which $\Phi_{\rm 1RSB}(m) = m \phi_{\rm RS}$ and one recovers the RS estimate; in the presence of clusters the minimizer is reached either in $m_{\rm s}(\ell)=1$ if $l_{\rm clust} < \ell < \ell_{\rm cond}$, or in some non-trivial value $m_{\rm s}(\ell)<1$ when $\ell > \ell_{\rm cond}$.

\item $\ell_{\rm r}$ is the smallest average degree for which the solution of the 1RSB equation at the static value $m_{\rm s}$ of the Parisi parameter has a positive fraction of hard fields. We will also denote $\ell_{\rm r}(m)$ the threshold defined similarly but at a fixed value of $m$. We will compute in particular $\ell_{\rm r}(m=1)$, and note that if $\ell_{\rm r}(m=1) < \ell_{\rm cond}$ then $\ell_{\rm r} = \ell_{\rm r}(m=1)$.

\end{itemize}

The numerical resolution of the cavity equations can be done via ``population dynamics'' algorithms (also known as particle representations in the context of filtering), that rely on the approximation of probability laws by the empirical measure of a large sample (population) of representative elements. For instance to solve the RS equation \eqref{eq:pdfmsg} one considers a sample of $\psi$'s, approximate $\Pm^{\rm RS}$ in the right hand side by the empirical distribution of the sample, and represent the left hand side as a new population of $\psi$'s. This procedure is repeated until convergence towards a fixed point solution of the equation, the numerical accuracy improving when the sample size increases. The same idea can be followed at the 1RSB level, with an additional difficulty: each $P$ in the sample representing $\Pm^{\rm{1RSB}}$ is itself a probability distribution, that has to be represented in an approximate way. This leads to a representation of $\Pm^{\rm{1RSB}}$ by a ``population of populations of $\psi$'s'', whose numerical accuracy is strongly limited by the memory available on current computers. Fortunately we have seen above that the determination of the thresholds $\ell_{\rm clust}$,  $\ell_{\rm cond}$ and $\ell_{\rm col}$ involves the resolution of the 1RSB equations only at $m=1$ and $m=0$. It turns out that these two specific values of $m$ allow great analytical simplifications, which we explain in the rest of this section before presenting our numerical results.

\subsubsection{Simplifications at $m=1$}

The special role played by the value $m=1$, as well as the connection between the 1RSB equations in this case and the tree reconstruction problem~\cite{Mossel04}, were first discussed in~\cite{MezardMontanari06,Krzakala2007,MoRiSe08,Zdeborova2007}, to which we refer the reader for more extensive discussions.

The first step in the simplification of \eqref{eq:pdfmsgmsg} consists in considering the normalization $\Z_1\itmu$ in \eqref{eq:nd1RSB}. At $m=1$, and thanks to the condition \eqref{eq:P_sym}, one sees that $\Z_1\itmu=\overline{\Z}_0\itmu$ where $\overline{\Z}_0\itmu$ is the value of \eqref{eq:defZ0} computed with $\psi\jtn=\overline{\psi}$ for all incoming messages. As the denominator $\Z_1\itmu$ is thus independent on the $\{P\jtn \}$ one realizes that the functional $\FRSB$ is linear in each of its argument. This allows us to introduce the average of $\Pm^{\rm{1RSB}}$,
\begin{align}
\overline{P} = \int {\rm d} P \, \Pm^{\rm{1RSB}}(P) \, P \ ,
\label{eq:def_overlineP}
\end{align}
and write a closed equation on it,
\begin{align}
\overline{P}(\psi)= &
\sum_{d=0}^{\infty} r_d 
\int \prod_{\nu = 1}^d \prod_{j=1}^{K-1} \mathrm{d} \psi\jtn \overline{P}(\psi\jtn) 
\notag \\ &
\kr{\psi}{\FRS(\br{\psi\jtn})} \frac{\Z_0(\{\psi\jtn\})}{\overline{\Z}_0} \; .
\label{eq:1RSBm1}
\end{align}
that easily follows from \eqref{eq:pdfmsgmsg}.

This is already a great simplification with respect to the general 1RSB formalism, as the unknown to solve for, $\overline{P}$, is simply a distribution over $\psi$'s instead of the distribution of distributions $\Pm^{\rm{1RSB}}$. There is still an annoying feature in this equation, namely the reweighting factor $\Z_0$, that we shall now get rid of.

To do this let us define, for all $x=1,\dots,q$, a biased version of $\overline{P}$ according to
\begin{align}
\overline{P}_x(\psi) = q \, \psi(x) \, \overline{P}(\psi) \ ,
\label{eq:def_oPx}
\end{align}
that we shall call a conditional version of $\overline{P}$. Each of these $q$ measures are normalized to 1 thanks to \eqref{eq:P_sym}, and can thus be interpreted as probability distributions. Reciprocally one can express $\overline{P}$ in terms of its conditional versions according to
\begin{align}
\overline{P}(\psi) = \frac{1}{q} \sum_{x=1}^q \overline{P}_x(\psi) \ .
\end{align}
Plugging these definitions into \eqref{eq:1RSBm1} leads to the following equations on the $\overline{P}_x$'s,
\begin{align}
\overline{P}_x^{(n+1)}(\psi)= 
\sum_{d=0}^{\infty} r_d 
\sum_{\mathbf{x}} & p(\mathbf{x}|x)
\int \prod_{\nu = 1}^d \prod_{j=1}^{K-1} \mathrm{d} \psi\jtn \overline{P}^{(n)}_{x_{\nu,j}}(\psi\jtn) 
\notag \\ &
\kr{\psi}{\FRS(\br{\psi\jtn})} \; ,
\label{eq:1RSBm1_cond}
\end{align}
where we have added for future convenience some indices $n+1$ and $n$ in the left and right hand side respectively, and where $\mathbf{x} = \{ x_{\nu,j}\}_{\nu=1,\dots,d}^{j=1,\dots,K-1}$, with
\begin{align}
p(\mathbf{x}|x) = \prod_{\nu = 1}^d \frac{1+(e^{-\beta}-1)\mathbb{I}(x=x_{\nu,1}=\dots=x_{\nu,K-1})}{q^{K-1}+e^{-\beta}-1} \ ;
\end{align}
here and in the following $\mathbb{I}(E)$ the indicator function of the event $E$. This is now a very convenient form for numerical resolution, as it only involves a finite number $q$ of probability distributions, without any reweighting factor. It is also sufficient to compute the thermodynamic properties of the model. Besides $\Phi^{\rm 1RSB}(m=1)$ which is simply equal to $\phi^{\rm RS}$ one needs the derivative of the replicated potential in order to deduce the complexity. This can be obtained by taking the derivative with respect to the explicit dependence in $m$ of \eqref{eq:phi1RSB}; expressing it in terms of the $\overline{P}_x$ yields
\begin{widetext}
\begin{eqnarray}
\left.\frac{\rm d}{{\rm d}m}\Phi^{\rm 1RSB}(m) \right|_{m=1} &=&  \sum_{d=0}^{\infty}  p_d \frac{1}{q}\sum_{x,\mathbf{x}}  p(\mathbf{x}|x) \int \prod_{\mu=1}^d\prod_{j=1}^{K-1} \mathrm{d} \psi\jtm \overline{P}_{x_{\mu,j}}(\psi\jtm) 
\Z_0^{i+\partial i} \ln \Z_0^{i+\partial i} \nonumber \\
&-& \frac{\ell(K-1)}{K} \sum_{x_1,\dots,x_K} p(x_1,\dots,x_K) \int \prod_{i=1}^K \mathrm{d} \psi_i \overline{P}_{x_i}(\psi_i) \Z_0^\mu \ln \Z_0^\mu
\; ,
\end{eqnarray}
\end{widetext}
where in the second term
\begin{align}
p(x_1,\dots,x_K) =  \frac{1+(e^{-\beta}-1)\mathbb{I}(x_1=\dots=x_K)}{q^K+q(e^{-\beta}-1)} \, .
\end{align}

In addition to its numerical convenience this formulation has an enlightening interpretation in terms of the tree reconstruction problem~\cite{Mossel04,MezardMontanari06} that we shall now describe. Consider indeed a Galton Watson hypertree with offspring distribution $r_d$, and draw at random a coloring of its vertices from the free boundary Gibbs measure at inverse temperature $\beta$. By this we mean the broadcasting procedure in which one draws the color $x$ of the root uniformly at random, then independently for each of the hyperedges around the root the colors of the $K-1$ other variables are chosen with probability proportional to $e^{-\beta}$ if the hyperedge is monochromatic, 1 otherwise, and so on and so forth for the next levels of the tree. Suppose now that the color of the vertices at distance $n$ from the root are revealed to an observer, whom is asked to guess the value of $x$ from this sole information. The best strategy the observer can follow is a Bayesian one, computing the marginal probability of the root conditional on the observations. As the problem is defined on a tree this computation can be done exactly with the Belief Propagation algorithm. A moment of though should convince the reader that the distribution of this marginal probability, conditional on the root being of color $x$ during the broadcast process, is nothing but the distribution $\overline{P}_x^{(n)}$ obtained after $n$ iterations of \eqref{eq:1RSBm1_cond} from the initial condition 
\begin{align}
\overline{P}_x^{(0)}(\psi) = \delta[\psi-\psi^x] \, .
\label{eq:HO_init}
\end{align}
Let us denote $\overline{P}_x = \lim_{n \to \infty} \overline{P}^{(n)}_x$ the fixed point solution of \eqref{eq:1RSBm1_cond} reached for infinitely deep trees. Two situations can arise: either $\overline{P}_x(\psi)=\delta(\psi-\overline{\psi})$, or the limit is non-trivial. In the first case one says that the tree problem is not reconstructible, as no information on the color of the root can be deduced from the observation of far away vertices, in the second it is reconstructible, as an observer can estimate the color of the root with a probability larger than from a random guess. As we have defined the clustering transition in terms of the existence of a non-trivial solution of the 1RSB equations at $m=1$, which we showed to be equivalent to \eqref{eq:1RSBm1_cond}, this non-reconstructible to reconstructible transition for the tree problem coincides with the clustering transition.

To quantify the distance between $\overline{P}_x$ and the trivial fixed-point we define the following overlap,
\begin{align}
C^{(n)} =
\frac{1}{1-1/q}
\int {\rm d} \psi \overline{P}_x^{(n)}(\psi) \sum_{x'=1}^q (\psi(x')-1/q)^2 \ ,
\label{eq:def_overlap}
\end{align}
which is independent of $x$ thanks to the symmetry between colors, and in which the prefactor ensures the normalization $C^{(0)}=0$. The reconstructibility of the tree problem, i.e. the condition $\ell > \ell_{\rm clust}$, is then equivalent to $\lim_{n \to \infty} C^{(n)} > 0$.

We shall now complete our study of the special case $m=1$ of the 1RSB formalism by studying the rigidity threshold $\ell_{\rm r}(m=1)$. We recall that we defined this transition in terms of the appearance of hard-fields in the 1RSB solution at zero temperature, see in particular \eqref{eq:def_eta}. We have thus to compute the probability of an hard-field $\psi^x$ under the distribution $\overline{P}$, that we shall denote $\eta$, or equivalently the probability $q \eta$ of $\psi^x$ under $\overline{P}_x$ (which does not contain hard-fields of colors $x'\neq x$, see \eqref{eq:def_oPx}). In terms of the tree reconstruction problem explained above, the question is now whether an observer can deduce the color of the root without any probability of error (rather than more accurately than a random guess) from the colors of far away vertices; this is sometimes called a naive reconstruction algorithm, as the observer only uses part of the information available by concentrating on the hard-fields.

To be more precise, let us call $q \eta^{(n)} = \overline{P}^{(n)}_x(\psi^x)$ and derive from \eqref{eq:1RSBm1_cond} a recursion relation of the form $\eta^{(n+1)}=f(\eta^{(n)})$. From \eqref{eq:defZ0} we see that $\FRS(\br{\psi\itm}) = \psi^x$ if and only if each of the $q-1$ colors $x'$ distinct from $x$ is forbidden by at least one adjacent constraint $\mu$; the latter event means that the $K-1$ other variables in $\mu$ are forced to the color $x'$, i.e. that $\psi\itm = \psi^{x'}$. For concreteness let us assume that $x=1$ and denote $d_{x'}$ the number of adjacent constraints forbidding the color $x'=2,\dots,q$, and $d_0=d-(d_2+\dots+d_q)$ the number of adjacent constraints that do not forbid any color. As the various constraints are independent in \eqref{eq:1RSBm1_cond} the vector $(d_0,d_2,\dots,d_q)$ has a multinomial distribution, the probability of one constraint $\mu$ forbidding one color $x'$ being $1/(q^{K-1}-1)$, the probability of $x_{\mu,1}=\dots=x_{\mu,K-1}$ under $p(\mathbf{x}|x)$, multiplied by $(q \eta^{(n)})^{K-1}$, the probability of picking only hard fields from the $\overline{P}_{x'}^{(n)}$ . Combining these observations leads easily to the following formula:
\begin{widetext}
\begin{align}
q \eta^{(n+1)} = \sum_{d=0}^\infty r_d \sum_{\substack{d_0,d_2,\dots,d_q \\ d_0+d_2+\dots+d_q =d}} \frac{d!}{d_0! d_2! \dots d_q!} 
\left( 1- \frac{(q-1)(q \eta^{(n)})^{K-1}}{q^{K-1}-1} \right)^{d_0}
\left(\frac{(q \eta^{(n)})^{K-1}}{q^{K-1}-1} \right)^{d_2+\dots+d_q}
\mathbb{I}(d_2 >0) \dots \mathbb{I}(d_q >0) 
\label{eq:rigidtym1}
\end{align}
\end{widetext}
To put this recursion under a more tractable form we use the following identity:
\begin{align}
\mathbb{I}(d_2 >0)& \dots \mathbb{I}(d_q >0) = (1-\delta_{d_2,0}) \dots (1-\delta_{d_q,0}) \nonumber \\ 
&= \sum_{p=0}^{q-1} (-1)^p \binom{q-1}{p} \delta_{d_2,0} \dots \delta_{d_{2+p-1},0} \ ,
\label{eq:incluexclu}
\end{align}
where we made a slight abuse of notation in the second line, symmetrizing all possible choices of $p$ colors. This identity is an analytical expression for the combinatorial inclusion-exclusion principle, expressing the conjunction of $q-1$ events in terms of conjuctions of the complementary events. Inserting \eqref{eq:incluexclu} in \eqref{eq:rigidtym1} one can perform the multinomial sums and obtain
\begin{equation}
\label{fpeqn_new}
  \eta^{(n+1)} = \frac{1}{q} \sum_{d=0}^\infty r_d\sum_{p=0}^{q-1}\, (-1)^p\, \binom{q-1}{p}\,
   \left(1 - \frac{p\, (q\eta^{(n)})^{K-1}}{q^{K-1}-1}\right)^d \, .  
\end{equation}
In the case of a Poissonian degree distribution of average $\ell$ this expression can be further simplified into
\begin{equation}
\eta^{(n+1)} = \frac{1}{q}  \left(1- \exp\left[-\ell \frac{(q\eta^{(n)})^{K-1}}{q^{K-1}-1}\right]\right)^{q-1} \ .
\label{eq:recursion_rigidity_m1}
\end{equation}
This recursion always admits the fixed-point solution $\eta=0$; the rigidity thrshold $\ell_{\rm r}(m=1)$ can be computed as the smallest $\ell$ such that a non-trivial fixed-point $\eta >0$ exists.

\subsubsection{Simplifications at $m=0$ and $\beta=\infty$}
\label{subsec:col}

We shall now explain the simplifications of the 1RSB cavity formalism that appears in the limit $m\to 0$, taken after the zero-temperature limit $\beta \to \infty$, and re-obtain the Survey Propagation (SP) equations first obtained for the $K$-SAT problem in~\cite{MezardParisi02}, then for the $q$-coloring for graphs in~\cite{Krzakala2004} (the order of the limits in $\beta$ and $m$ is slightly different in these papers but the final result is the same).

\paragraph{SP equations and complexity}

The first point to notice is that the Parisi parameter $m$ always appear in quantities of the form $\Z_0^m$, that can be thus replaced in the $m\to 0$ limit by $\mathbb{I}(\Z_0>0)$. The probability that $\Z_0$ vanishes turns out to depend only on the intensity of hard-fields in the 1RSB distributions, which is the key to the simplifications in the SP formalism.

More precisely, let us first focus on the expression of $\Z_0\itmu$ given in \eqref{eq:defZ0}. This quantity vanishes if and only if each of the $q$ colors $x$ is forbidden by at least one constraint around it, which in turns means that the $K-1$ other variables are forced by an hard-field to be of color $x$. With the same kind of reasoning as the one we did to arrive at (\ref{eq:rigidtym1}-\ref{fpeqn_new}), we see that the normalization $\Z_1\itmu$ of equation \eqref{eq:nd1RSB} becomes in the $m=0$ limit:
\begin{align}
\Z_1\itmu = \sum_{p=1}^{q} (-1)^{p+1} \binom{q}{p} \prod\nbe{\nu}{i}{\mu}  \p{1 - p \prod\nbe{j}{\nu}{i} \n\jtn}
\end{align}
Similarly the integral in \eqref{eq:nd1RSB} will give rise to $\delta[\psi\itm - \psi^x]$ if the $q-1$ colors distinct from $x$ are forbidden while $x$ is not (the ill-defined situation with the $q$ colors forbidden is cancelled out by the factor $\Z_0$ in the integral). Putting together these two observations allows us to project \eqref{eq:nd1RSB} onto the intensity of hard-fields alone, that obey the recursion:
\begin{align}
\label{eq:SPrec}
\n\itm &= \frac{\displaystyle\sum_{p=0}^{q-1} (-1)^p \dbinom{q-1}{p} 
          \prod\nbe{\nu}{i}{\mu}  \p{1 - \p{p+1} \prod\nbe{j}{\nu}{i} \n\jtn} }
      {\displaystyle\sum_{p=1}^{q} (-1)^{p+1} \dbinom{q}{p} 
          \prod\nbe{\nu}{i}{\mu}  \p{1 - p \prod\nbe{j}{\nu}{i} \n\jtn} } 
\nonumber \\
      &\equiv \FSP(\{\n\jtn\}) \, ,
\end{align} 

The probability distribution $\Pm^{\rm 1RSB}(P)$ of the generic 1RSB treatment of random hypergraph ensembles can be simplified into a distribution $\Pm^{\rm{SP}}(\n)$ for the intensity of hard-fields. From \eqref{eq:pdfmsgmsg} one obtains easily the fixed-point equation it obeys, namely
\begin{align}
\label{eq:pdfSPmsg}
\Pm^{\rm{SP}}(\n) =  \sum_{d=0}^{\infty} r_d & \int \prod_{\nu = 1}^d \prod_{j=1}^{K-1} \mathrm{d} \n\jtn 
\Pm^{\rm{SP}}(\n\jtn) \notag \\
&\kr{\n}{\FSP(\br{\n\jtn})} \;.
\end{align}

The $m=0$ complexity, whose vanishing will yield the colorability threshold, can be computed from the solution of this equation. From \eqref{eq:SigmaLegendre} we see that $\Sigma(m=0)=\Phi^{\rm 1RSB}(m=0)$, because when $m=0$ all the clusters are counted in the same way, irrespectively of their sizes. Taking the limit $m\to 0$ in the expression \eqref{eq:phi1RSB} of $\Phi^{\rm 1RSB}$ gives
\begin{align}
\label{eq:SPcomplexity}
\Sigma&(m=0) =  
\sum_{d=0}^\infty p_d \int \prod_{\mu = 1}^d \prod_{j=1}^{K-1} 
\mathrm{d} \n\jtm \Pm^{\rm{SP}}(\n\jtm) \ln \Z_{\rm SP}^{i + \partial i} \notag\\
&- \frac{\ell (K-1)}{K} \int \prod_{i=1}^K \mathrm{d} \n_i \Pm^{\rm{SP}}(\n_i) \ln \Z_{\rm SP}^{\mu} \ ,
\end{align}
where the expressions of $\Z_{\rm SP}$ are obtained from (\ref{eq:defZ1idi},\ref{eq:defZ1mu}) in the limit $m\to 0$. Analyzing once more the probabilities that $\Z_0^{i + \partial i}$ and $\Z_0^{\mu}$ does not vanish because of contradicting hard-fields yields
\begin{align}
\Z_{\rm SP}^{i+\partial i}&(\br{\n\jtm}) = \notag \\
&\sum_{p=1}^{q} (-1)^{p+1} \dbinom{q}{p} \prod_{\mu=1}^d \p{1 - p \prod_{j=1}^{K-1} \n\jtm} \; ,
\end{align}
which is one minus the probability that the $q$ colors are forbidden by neighboring constraints, and
\begin{align}  
\Z_{\rm SP}^{\mu} (\n_1,\dots,\n_K) = 1 - q \prod_{i=1}^K \n_i \; ,
\end{align}
where one recognizes in $\prod \n_i$ the probability that an hyperedge is forced to be monochromatic of one given color.

\paragraph{Stability analysis}
 
As its name suggests the 1RSB version of the cavity method is only the first level of a hierarchy of refinements. In the $p$-RSB version there are $p$ steps of replica symmetry breaking, which means that the pure states (clusters) introduced at the 1RSB level are themselves organized in a hierarchical structure with $p$ levels. The equations at the $p$-th level admit as special cases the solutions of the equations at all the $p'<p$ levels (we saw above that the RS formalism was recovered as a trivial solution of the 1RSB equations). In principle the correct (in the sense of the cavity method) description of a mean-field model should be sought for in all the levels of RSB; in simple situations this hierarchy is expected to collapse on the RS and 1RSB level only, i.e. in such cases the $p$-RSB equations do not admit solutions that do not reduce to the RS or 1RSB ones. Solving the cavity equations with an arbitray level of RSB in a sparse mean-field model is an extremely challenging task, hence this program cannot be implemented in practice. One can however test the local stability of the 1RSB solution into the larger set of 2RSB solutions, which allows to assess the existence, or not, of nearby proper 2RSB solutions which presumably strictly improve the 1RSB bound \eqref{eq:min_Phi1RSB}. This local stability analysis of the 1RSB solutions for sparse models was introduced in~\cite{Montanari2003,MontanariParisi04}, see also~\cite{Mertens2005,Krzakala2004,RivoireBiroli04} for other presentations of the method. The 2RSB description of the set of solutions of a CSP introduces clusters of configurations, which are themselves organized into groups of clusters. As a consequence there are two ways in which this can be reduced to the 1RSB description: either there is a single group, or each cluster contains a single configuration. These two reductions yield two different type of instabilities of the 1RSB solution, termed respectively type I or noise propagation instability, and type II or bug proliferation instability. To quantify these phenomena one introduces two positive numbers $\lambda_{\rm I}$ and $\lambda_{\rm II}$, in such a way that one type of perturbation of the 1RSB solution is stable (resp. unstable) if the corresponding $\lambda$ is strictly smaller (resp. larger) than 1. One finds generically that $\lambda_{\rm I}$ grows with the density of constraints $\ell$, crossing the critical value 1 at a threshold denoted $\ell^{\rm SP}_{\rm I}$, the 1RSB solution being thus stable for $\ell < \ell^{\rm SP}_{\rm I}$. Conversely $\lambda_{\rm II}$ is usually found to be decreasing with $\ell$, the stability regime being thus of the form $\ell > \ell^{\rm SP}_{\rm II}$. We defer the details of the computations of the $\lambda$ parameters for the hypergraph coloring problem to Appendix \ref{app:SP}, the numerical results being presented in the next Section.

\section{Results}
\label{sec:results}
We present in this Section the results obtained on the coloring of random hypergraphs through the application of the cavity method presented above. We shall first give the numerical values of the various thresholds of the Erd\H{o}s-R\'enyi ensemble in Sec.~\ref{sec:results_ER}, then study briefly a regular case (cf. Sec.~\ref{sec:results_regular}), and finally present analytical asymptotic expansions of the thresholds for large values of $K$ and/or $q$ in Sec.~\ref{sec:asymptotic}.

\subsection{The thresholds of the Erd\H{o}s-R\'enyi ensembles}
\label{sec:results_ER}
Our main results are summarized in Table~\ref{tab:results} in which we give, for a few small values of $K \ge 3$ (the graph case $K=2$ has already been studied in~\cite{Krzakala2004,Zdeborova2007}) and $q$, the average degree $\ell$ at which the various phase transitions occur in the Erd\H{o}s-R\'enyi ensemble, i.e. when the degree and excess degree distributions $p_d$ and $r_d$ are Poissonian laws of average $\ell$.

\begin{table*}
\begin{center}
\begin{tabular}{p{0.18\linewidth}
p{0.07\linewidth}
p{0.09\linewidth}
p{0.07\linewidth}
p{0.07\linewidth}
p{0.07\linewidth}
p{0.07\linewidth}
p{0.07\linewidth}
p{0.07\linewidth}
p{0.07\linewidth}}
\hline
& $\ell_{\rm clust}$ 
& $\ell_r(m=1)$
& $\ell_{\rm cond}$
& $\ell_{\rm col}$
& $\ell_{\rm RS}$
& $\ell^{\rm RS}_{\rm stab}$
&  $\ell^{\rm SP}_{\rm I}$
& $\ell^{\rm SP}_{\rm II}$\\
\hline\hline \\
$q= 2 $ $K= 3 $ 
    & 4.50
    & (7.37)
    & 4.50
    & (6.32)
    & 7.23
    & {\bf 4.50}
    & {\bf 6.07}
    & 6.16
\\
$q= 2 $ $K= 4 $ 
    & 16.33 
    & (21.62)
    & 16.33
    & (19.62)
    & 20.76
    & {\bf 16.33}
    & {\bf 19.35}
    & 17.84
\\
$q= 2 $ $K= 5 $ 
    & 47.4
    & (52.63)
    & 51.5
    & 52.32
    & 53.70
    & 56.25
    & 59.42
    & 43.9
\\
$q= 3 $ $K= 3 $ 
    & 25.06 
    & (28.07)
    & 26.2
    & 26.92
    & 27.98
    & 32.00
    & 33.62
    & 23.9
\\
$q= 3 $ $K= 4 $ 
    & 97.7
    & 105.88
    & 114.3
    & 115.04
    & 116.44
    & 225.33
    & 225.51
    & 90.7
\\
$q= 4 $ $K= 3 $ 
    & 56.20
    & 61.09
    & 62.7
    & 63.3
    & 64.44
    & 112.50
    & 112.78
    & 52.7
\\
\end{tabular}
\end{center}
\caption{The thresholds of $q$-coloring of $K$-uniform Erd\H{o}s-R\'enyi random hypergraphs, given in terms of their average degree $\ell$. The definitions of the thresholds are given in the text and summarized in Table \ref{tab:thresholds}. Bold font numbers indicate significant instabilities. Numbers given by unstable or invalid ansatz are put between parenthesis: (i) The prediction of the colorability threshold for the two first rows is hindered by the SP type I instability. (ii) The $m=1$ rigidity threshold does not coincide with the true rigidity threshold for the four first rows because $\ell_{\rm cond} < \ell_{\rm r}(m=1)$: the rigidity should be evaluated at the static Parisi parameter $m_{\rm s}<1$ in those cases.}
\label{tab:results}
\end{table*}

Some of these thresholds have been obtained analytically or by solving numerically a simple scalar equation: the upperbound $\ell_{\rm RS}$ on the colorability threshold and the Kesten-Stigum bound $\ell^{\rm RS}_{\rm stab}$ can be read directly from \eqref{eq:lRS} and \eqref{eq:KSbound}, and the rigidity threshold $\ell_{\rm r}(m=1)$ is the smallest $\ell$ such that \eqref{eq:recursion_rigidity_m1} admits a positive fixed-point solution. 

The determination of the other thresholds have required an heavier numerical work. As explained in Sec.~\ref{sec:thresholds_computation} they are defined in terms of the solution of the 1RSB equations for the values $m=1$ and $m=0$ of the Parisi parameter. For these two special cases we have derived above some fixed-point equations where the unknowns are probability distributions of scalars (or $q$-dimensional vectors), see in particular \eqref{eq:1RSBm1_cond} and \eqref{eq:pdfSPmsg}. These equations can thus be handled numerically by population dynamics algorithms where the unknown probability distributions are approximated by the empirical distribution of large samples of random representants, that are iteratively updated until numerical convergence (see for instances the appendices of~\cite{Zdeborova2009} for more details on algorithmic implementation issues).

\begin{figure*}
\begin{center}
      \includegraphics[width=\textwidth]{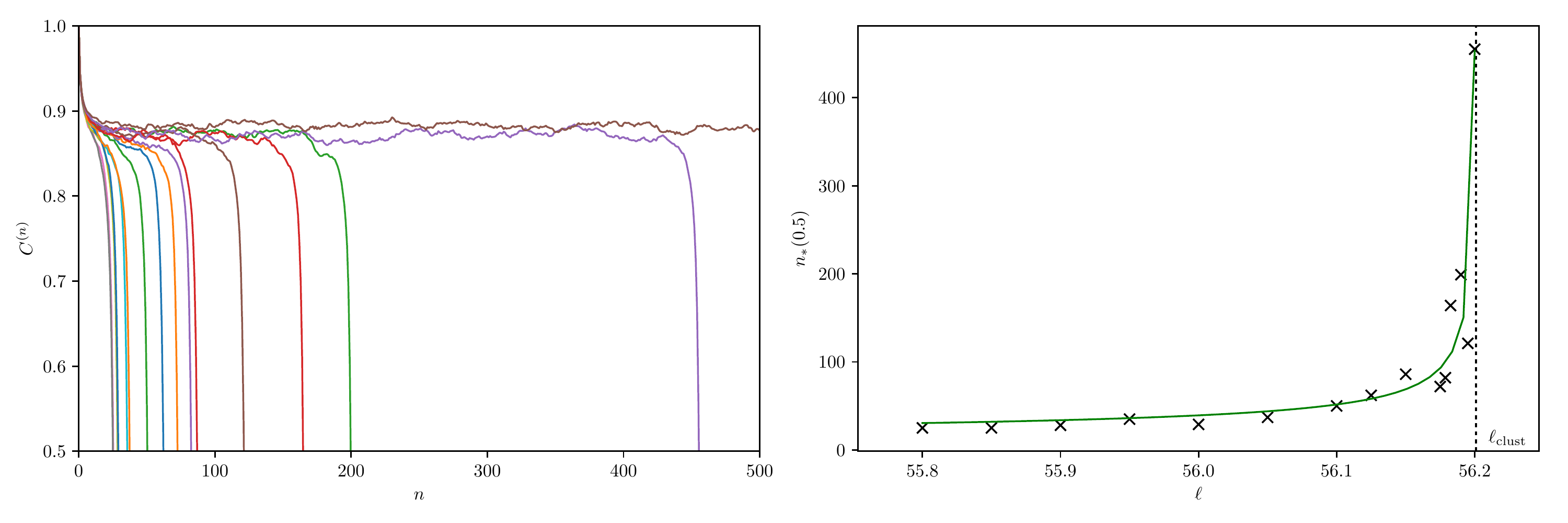}
     \caption{Left: the overlap (or point-to-set correlation function) $C^{(n)}$ of equation \eqref{eq:def_overlap} as a function of the number $n$ of iterations, for the $q=4$-coloring of $K=3$-uniform ER hypergraphs. The different curves corresponds to the following values of the average degree $\ell$, from left to right : 55.8, 55.85, 55.9, 55.95, 56.0, 56.05, 56.1, 56.125, 56.15, 56.175, 56.1788, 56.1825, 56.19, 56.195, 56.2, 56.3. This last value corresponds to $\ell > \ell_{\rm clust}$, hence the overlap does not decay to 0. Right: the divergence of the number $n_*(0.5)$ of iterations after which the overlap drops below $0.5$ (black crosses), along with a fit of the form $A \times (\ell - \ell_{\rm clust} )^{-1/2}$ (green line) over the adjustable parameters $A$ and $\ell_{\rm clust}$, which yields the estimate $\ell_{\rm clust} \approx 56.2$ reported in Table~\ref{tab:results}. 
\label{fig:q4K3fitconvtime}}
\end{center}
\end{figure*}

\begin{figure*}
\begin{center}
      \includegraphics[width=\textwidth]{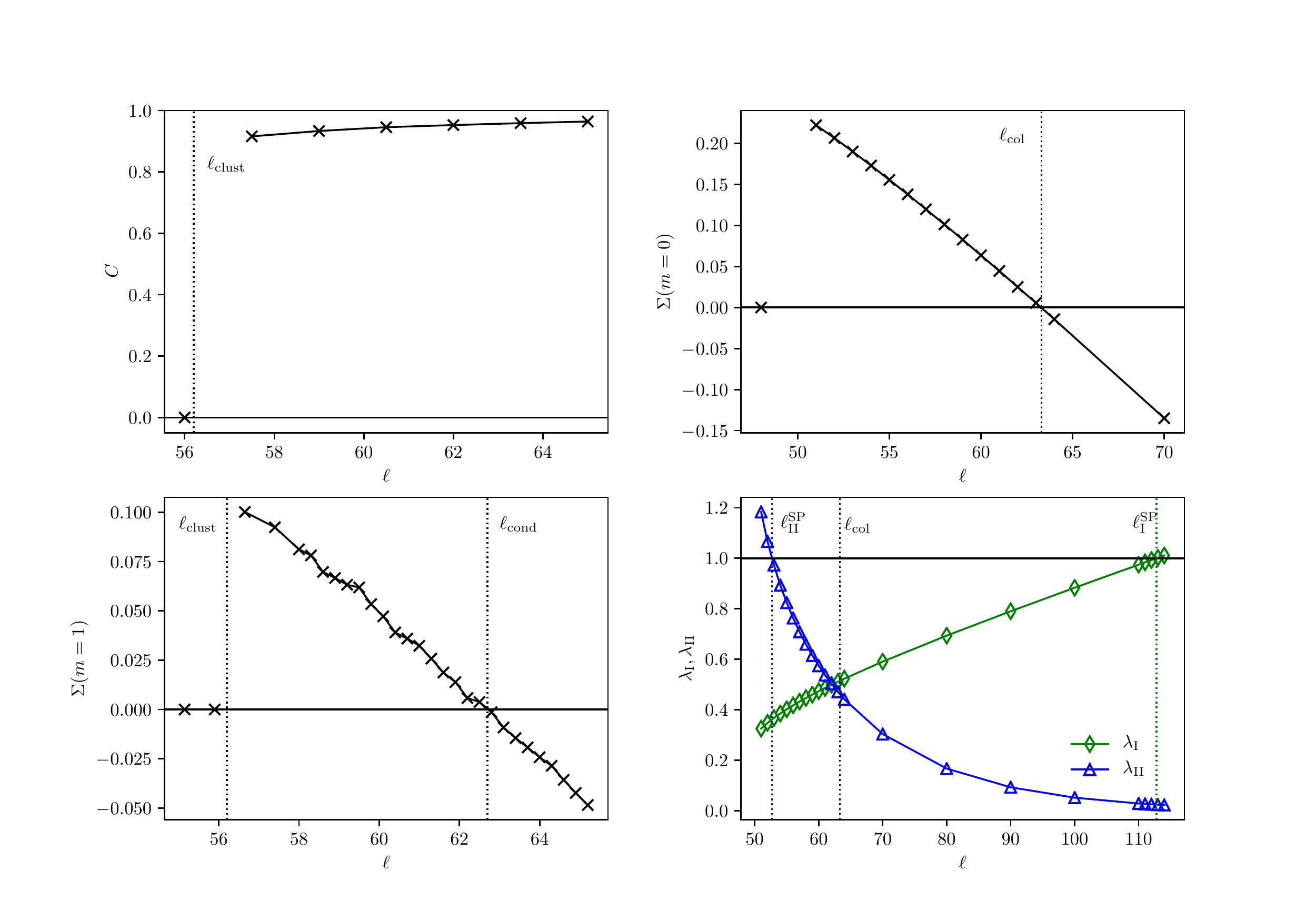}
     \caption{Numerical results for the $q=4$-coloring of $K=3$-uniform ER random hypergraphs. Left: Overlap (top) and complexity (bottom) for the solution of the 1RSB equation at $m=1$ as a function of the average degree $\ell$. Right: Complexity (top) and stability criteria (bottom) for the solution at $m=0$ (Survey Propagation) as a function of the average degree $\ell$. The stability of the SP solution is assessed by the quantities $\lambda_{\rm I}$ and $\lambda_{\rm II}$ defined in Appendix \ref{app:SP}, unstable regimes corresponding to $\lambda >1$.
\label{fig:m0}}
\end{center}
\end{figure*}

To be more specific we shall first consider the last line of Table~\ref{tab:results}, i.e. the case $q=4$, $K=3$, and explain the steps that led us to the numbers displayed in the Table. On the left panel of Figure~\ref{fig:q4K3fitconvtime} we plot the overlap $C^{(n)}$ (also known as the point-to-set function) defined in Eq.~\eqref{eq:def_overlap}, as a function of the number of iterations $n$, for several values of $\ell$. By definition the clustering threshold is the smallest value of $\ell$ such that this function does not decay to zero at large $n$, as it marks the appearance of a non-trivial solution of the 1RSB equations at $m=1$. One can clearly see on this plot that the bifurcation is here discontinuous: the large $n$ limit of $C^{(n)}$ jumps from zero to a strictly positive value when $\ell$ crosses $\ell_{\rm clust}$. Accordingly we have here $\ell_{\rm clust} < \ell_{\rm stab}^{\rm RS}$: this transition is unrelated to the Kesten-Stigum local instability of the trivial solution. A precursor of the transition when $\ell$ approaches $\ell_{\rm clust}$ from below is the divergence of the length of the ``plateau'' in $C^{(n)}$, that we quantify by $n_*(C)$, the number of iterations necessary to make the correlation drop below a specified level $C$. As illustrated on the right panel of Fig.~\ref{fig:q4K3fitconvtime} we have thus estimated $\ell_{\rm clust}$ by fitting the divergence of $n_*(C)$ as $(\ell - \ell_{\rm clust} )^{-1/2}$ close to the transition (see~\cite{MontanariSemerjian06} for a justification of this critical exponent).
We turn now our attention to Figure~\ref{fig:m0}. On its left part we continue the presentation of the results at $m=1$ initiated in Figure~\ref{fig:q4K3fitconvtime}; the upper left panel displays indeed the large $n$ limit of the overlap $C^{(n)}$ for $\ell > \ell_{\rm clust}$. The bottom left panel shows the value of the associated complexity $\Sigma(m=1)$: it is strictly positive at the birth of the non-trivial solution, and decreases with increasing $\ell$. The condensation threshold $\ell_{\rm cond}$ is deduced from this data as the point where the complexity vanishes. The right part of Figure~\ref{fig:m0} is devoted to the results of the Survey Propagation equations (i.e. the 1RSB equations at $m=0$). On the upper right panel we plotted the complexity $\Sigma(m=0)$, whose vanishing marks the colorability transition $\ell_{\rm col}$, while the lower right panel demonstrates that this prediction falls into a regime where the SP solution is locally stable, according to the analysis presented in Appendix \ref{app:SP}.

Let us underline the main qualitative features that arise from the analysis of the data we just presented in the case $q=4$, $K=3$: (i) the clustering transition occurs discontinuously, hence $\ell_{\rm clust} < \ell_{\rm stab}^{\rm RS}$ (ii) at this point the $m=1$ complexity is strictly positive, yielding $\ell_{\rm clust} < \ell_{\rm cond}$ (iii) the SP formalism is stable at $\ell = \ell_{\rm col}$, we have thus no reason to discard this prediction as a conjecture for the exact colorability threshold. We found the same qualitative features in the analysis of the cases $(q,K)=(2,5),(3,3),(3,4)$, see the Table for the numerical values which are of course quantitatively different, and we expect this to be the generic scenario for all larger values of $q$ and $K$ (i.e. for $q=2$ and $K\geq5$, and for $q\geq3$ and $K\geq 3$).

There are however two cases which do not correspond to this generic scenario, namely the bicoloring ($q=2$) of $K=3$ and $K=4$-uniform hypergraphs, see the first two lines of Table~\ref{tab:results}. The first aspect in which they differ is that the SP formalism is unstable (with respect to the noise propagation instability) for $\ell=\ell_{\rm col}$: the 1RSB computation does not give the correct prediction for the colorability threshold in such a situation, however one expects on the basis of the bounds of~\cite{Guerra2003,FranzLeone03,PanchenkoTalagrand04} the 1RSB result to be an upperbound on the location of the colorability transition (in fact one expects this bound to be strict, assuming that the instability implies the existence of a proper 2RSB solution that strictly improves the 1RSB bound of \eqref{eq:min_Phi1RSB}).

\begin{figure*}
\begin{center}
\includegraphics[width=\textwidth]{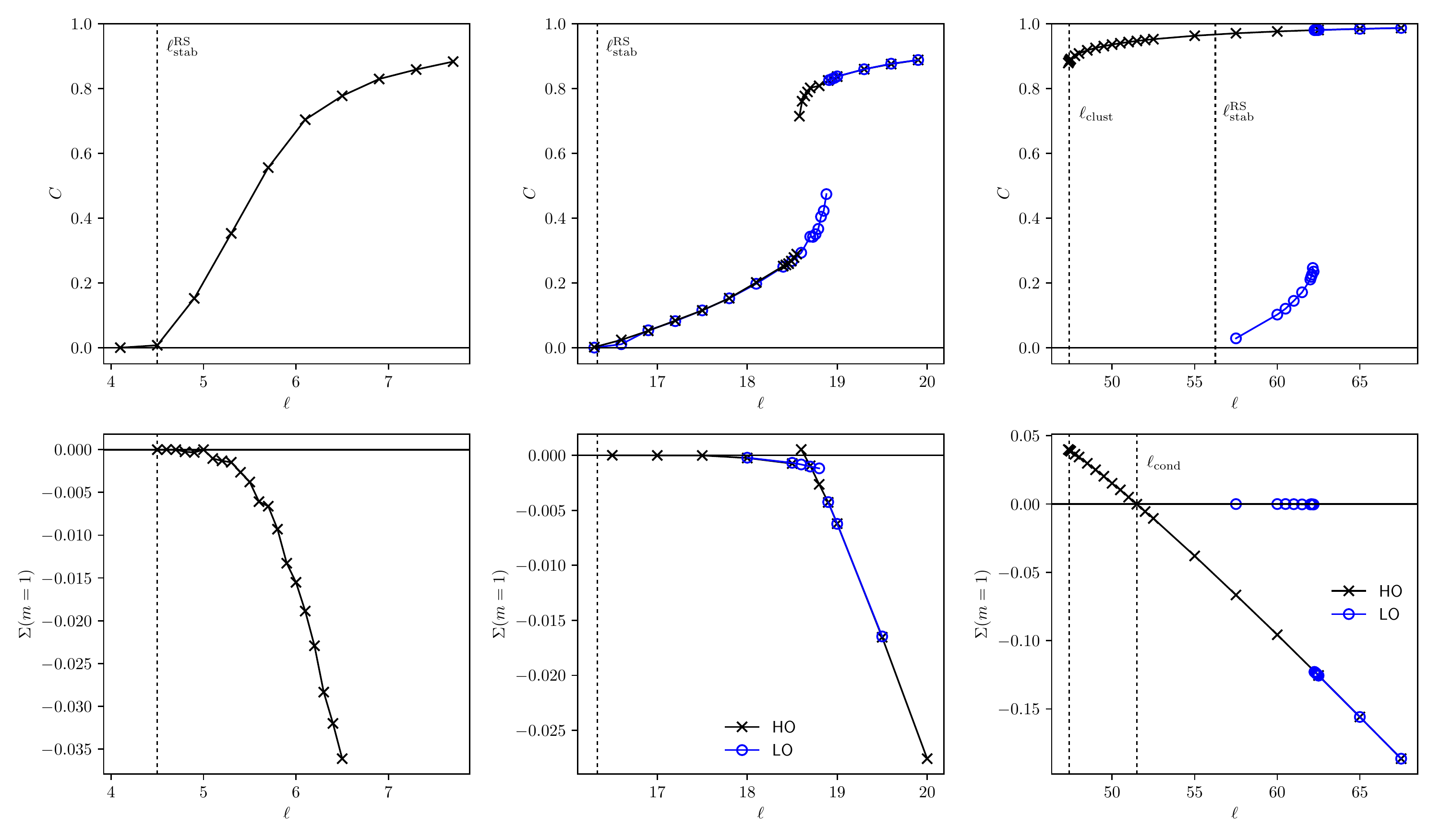}
\caption{Overlaps (top) and complexities (bottom) for the solutions of the 1RSB equations at $m=1$ as a function of the average connectivity $\ell$ of the ER random hypergraph ensemble, for $q=2$, $K=3$ (left), $K=4$ (middle) and $K=5$ (right). In the middle and right panels the two branches are selected by different initial conditions, see the text for details. 
\label{fig:m1}}
\end{center}
\end{figure*}

They also have a different pattern of transitions in the solution of their 1RSB equations at $m=1$. To demonstrate this more clearly we present in Figure~\ref{fig:m1} the overlap and complexity at $m=1$ for the bicoloring ($q=2$) of $K=3,4$ and $5$-uniform ER hypergraphs.  The case $K=3$ on the left exhibits a continuous transition, with $\ell_{\rm clust} = \ell_{\rm cond} = \ell_{\rm stab}^{\rm RS}$. The non-trivial solution of the 1RSB equation bifurcates continuously from the trivial one at the Kesten-Stigum threshold, its complexity is always negative hence there is no clustered un-condensed phase; this scenario was also found in the 3-coloring of ER graphs ($K=2$, $q=3$) in~\cite{Zdeborova2007}.

\begin{figure}
\includegraphics[width=0.4\textwidth]{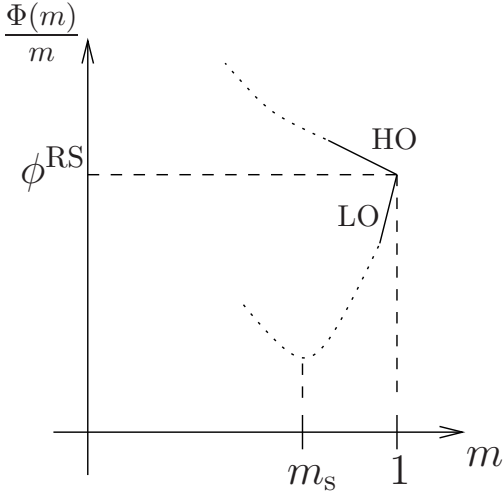}
\caption{A sketch of the behavior of $\Phi(m)/m$ when two solutions of the 1RSB equations coexist at $m=1$, the one denoted LO (resp. HO) has a negative (resp. positive) complexity, which fixes the sign of the slope in $m=1$. The dotted part of the curves are educated guesses, to be confirmed in Sec.~\ref{sec:results_regular}}.
\label{fig:sketch_Phi}
\end{figure}

The middle panel presents the rather peculiar results of the $K=4$ case. One sees the overlap growing continuously for $\ell >\ell^{\rm RS}_{\rm stab}=\ell_{\rm clust}$, as in the $K=3$ case, but it undergoes an upward jump for a larger value of the average degree. More precisely there is a domain of coexistence of two distinct stable solutions of the $m=1$ 1RSB equations, for $\ell$ roughly between $18.5$ and $19$. We call these two solutions the high and low overlap (HO/LO) branches, with two spinodals at which one of these branches disappear, both spinodals occuring for some $\ell > \ell_{\rm clust}$. In this domain of coexistence one obtains the HO branch by initializing the resolution of \eqref{eq:1RSBm1_cond} with the initial condition \eqref{eq:HO_init}, corresponing to the usual tree reconstruction problem, while the LO solution is reached from an initialization very close to the unstable trivial fixed-point $\delta[\psi -\overline{\psi}]$ (that we perturb by a small fraction of hard-fields), that implements the so-called robust version of the reconstruction problem~\cite{JansonMossel04}. One can see on the middle bottom panel of Fig.~\ref{fig:m1} that the $m=1$ complexity  of the LO branch is always negative, while a small part of the HO branch, for $\ell$ slightly above its spinodal, is positive. This phenomenon is puzzling at first sight, and its interpretation requires a moment of thought. One could think that in this regime the HO branch is the relevant one, hence that a small clustered un-condensed regime lies between two condensed ones. It turns out that this interpretation is wrong: in the variational prescription for the computation of the free-entropy recalled in~\eqref{eq:min_Phi1RSB} one should also minimize over the different solutions of the cavity equations in case several do coexist (the rigorous bounds of~\cite{Guerra2003,FranzLeone03,PanchenkoTalagrand04} actually hold for arbitrary order parameters, not necessarily solutions of the cavity equations). It is useful at this point to take a look at the sketch of the function $\Phi(m)/m$ presented in Figure~\ref{fig:sketch_Phi}, recalling that the slope of this function is given in terms of the complexity by $\frac{\rm d}{{\rm d}m} (\Phi(m)/m)= -\Sigma(m)/m^2$. This should convince the reader that the existence of one solution of the 1RSB equations with $\Sigma(m=1)<0$ implies that $\phi < \phi_{\rm RS}$ and that the model is in a condensed phase, even if other solutions have a positive complexity. The minimum of $\Phi(m)/m$ occurs indeed at a non-trivial static Parisi parameter $m_{\rm s}<1$, and most probably on the branch continuously connected to the solution with the most negative complexity at $m=1$ (if one excludes additional crossings of the branches as functions of $m$). The conclusion of this analysis is that $\ell_{\rm clust}=\ell_{\rm cond}=\ell^{\rm RS}_{\rm stab}$ in the case $q=2$, $K=4$, and that the coexistence of solutions of the 1RSB equations at $m=1$ is not thermodynamically relevant here: the minimum in~\eqref{eq:min_Phi1RSB} is reached at a non-trivial value $m_{\rm s}<1$, hence neither of the solutions at $m=1$ yields the correct 1RSB prediction for the entropy. Note that this peculiar bifurcation scenario remained, as far as we know, unobserved previously, and that it caused a slight mistake in~\cite{BrDaSeZd16}.

Finally the right panel of Figure~\ref{fig:m1} shows that there are also coexisting 1RSB solutions for $K=5$, but at variance with the $K=4$ case the spinodal of the HO branch occurs before the Kesten-Stigum transition, as well as the vanishing of the associated complexity. Hence in this case $\ell_{\rm clust} < \ell_{\rm cond} < \ell^{\rm RS}_{\rm stab}$: from a thermodynamics point of view we are in the generic case explained above for $q=4$, $K=3$, and the existence of the LO branch bifurcating continuously from the trivial fixed point at the Kesten-Stigum transition is completely irrelevant.

Let us briefly mention that the coexistence of solutions has much more important consequences in so-called planted ensembles of inference problems, which will be discussed in~\cite{RiSeZd17}. In this perspective determining the conditions for the existence of a continuously bifurcating solution above the Kesten-Stigum transition is an important question; it is shown in~\cite{RiSeZd17} that a large class of $K$-wise interacting Ising spin models (including the bicoloring of hypergraphs) always exhibit this continuous transition, as confirmed by the plots of Fig.~\ref{fig:m1}. Moreover the analysis of dense inference problems that mimick the large degree limit of the $q$-coloring of $K$-uniform hypergraphs~\cite{Lesieur2017} show that the criterion for the existence (resp. absence) of the continuous solution is $qK-2K-q < 0$ (resp. $qK-2K-q > 0$). This is agreement with the previous statement for $q=2$, and implies that when $K \ge 3$ and $q \ge 3$ there should not be a stable solution bifurcating continuously above Kesten-Stigum (except possibly in the marginal case $q=K=3$ which cannot be decided from the above criterion). Our numerical simulations seem to confirm this statement (and to indicate the absence of continuous solution when $q=K=3$ for Poisson hypergraphs), however we cannot formally exclude from them the existence of a very short branch with a spinodal closely above the Kesten-Stigum transition.

\subsection{A finite $m$ study of the bicoloring of regular hypergraphs ($q=2$, $K=4$)}
\label{sec:results_regular}
\begin{figure*}
\begin{center}
\includegraphics[width=\textwidth]{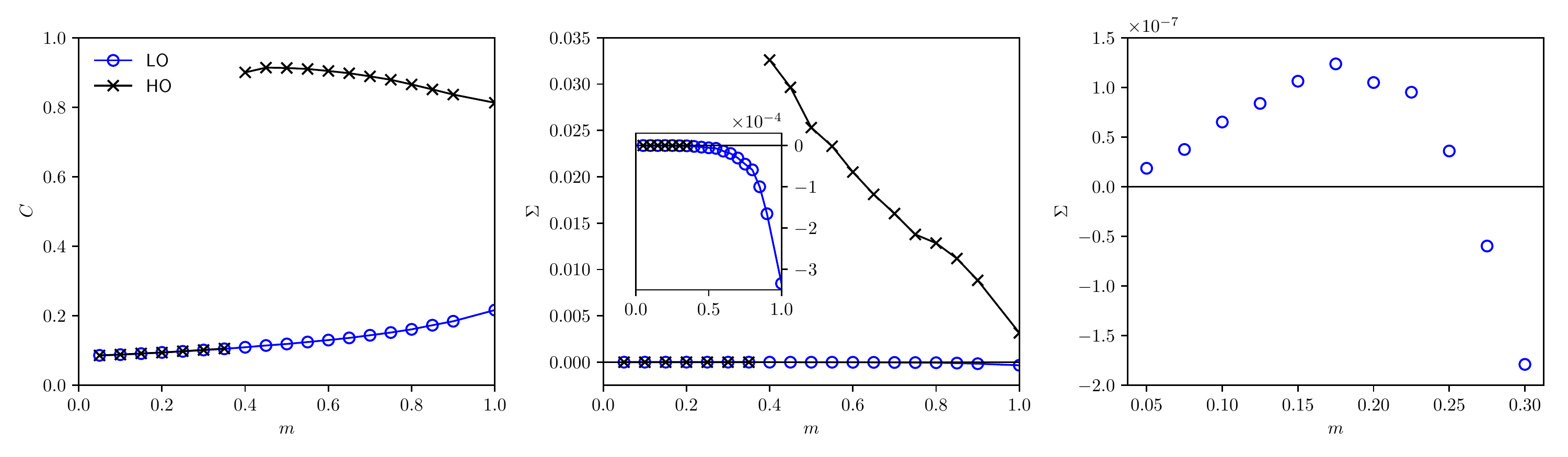}
\caption{Overlap (left) and complexity (middle, inset and right) of the solutions of the 1RSB cavity equations for the bicoloring of 19-regular 4-uniform random hypergraphs at temperature $T=0.1$, as a function of the Parisi parameter $m$. Inset of middle panel and right panel are zoomed views of middle pannel. Black lines with crosses corresponds to the high overlap branch (HO) and blues lines with blue circles to the low overlap (LO) branch, reached from the two different initial conditions explained in Sec.~\ref{sec:results_ER}. 
\label{fig:regT01}}
\end{center}
\end{figure*}

\begin{figure}
\begin{center}
\includegraphics[width=0.5\textwidth]{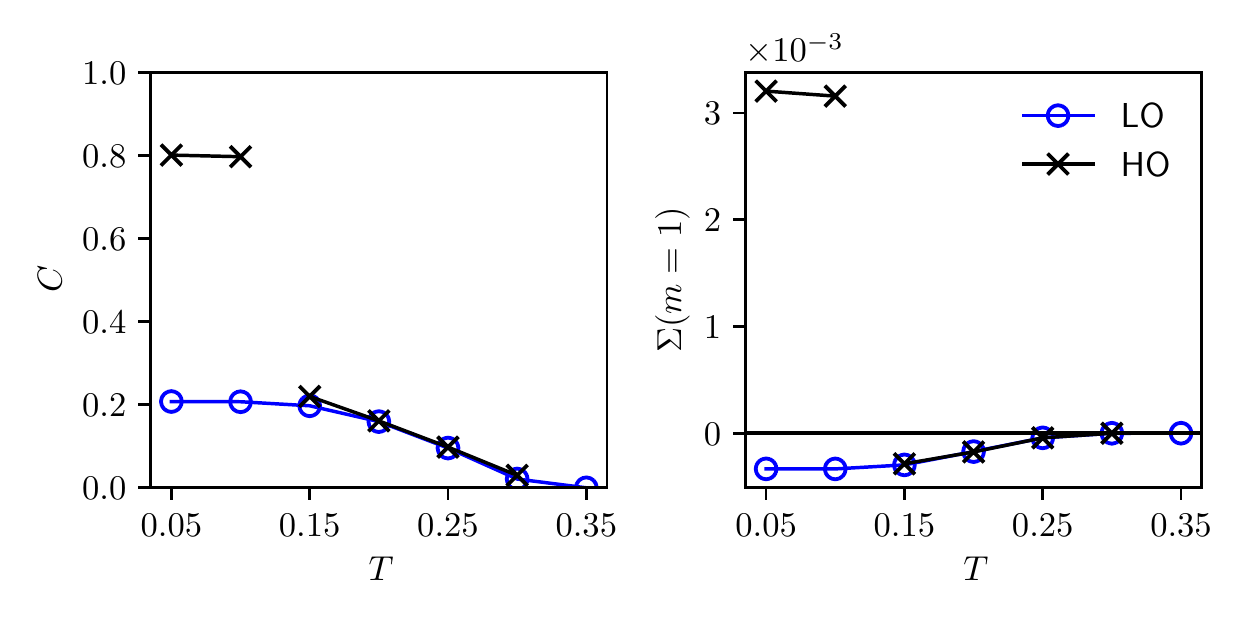}
\caption{Overlap (left) and complexity (right) of the solutions of the 1RSB cavity equations at $m=1$, for the bicoloring of 19-regular 4-uniform random hypergraphs as a function of the temperature. Same symbols as in Fig.~\ref{fig:regT01}.
\label{fig:regm1}}
\end{center}
\end{figure}

In order to confirm our interpretation of the coexisting solutions we found in the Erd\H{o}s-R\'enyi ensemble with $q=2$, $K=4$ and average degree between 18.5 and 19 (roughly) we should solve the 1RSB equations for arbitrary values of $m$ in $[0,1]$ to find explicitly the behavior sketched in Fig.~\ref{fig:sketch_Phi}. Unfortunately this is a rather difficult numerical task: apart from the two special cases $m=0$ and $m=1$ the solution of the 1RSB equations for arbitrary $m$ and non-trivial degree distributions have to be represented by populations of populations, a precise computation of $\Phi(m)$ is thus extremely challenging. To circumvent this difficulty we turned instead to regular ensembles, with deterministic degree distributions $p_d=\delta_{d,\ell}$ and $r_d=\delta_{d,\ell-1}$. In such a case the cavity equations presented in Sec.~\ref{subsec:1RSBcavity} admit a ``translationally invariant'' (sometimes called factorized) solution with
\begin{align}
\Pm^{\rm{1RSB}}(P) = \delta[P- P_{\rm reg}] \ ,
\end{align}
traducing the equivalence of all local neighborhoods in a regular random graph. The 1RSB equation with arbitrary $m$ can thus be handled with a single population of messages $\Y$ representing the distribution $P_{\rm reg}$ fixed-point solution of~\eqref{eq:closedmsgmsg}. Solving this equation is slightly more difficult than, for instance, \eqref{eq:pdfSPmsg}, because of the reweighting factor $\Z_0^m$. This point can be tackled by representing $P_{\rm reg}$ as a weighted sample of representants, or by using resampling techniques, we refer the reader to appendix E3 of~\cite{Zdeborova2009} for more details on the implementation issues.

We performed this study for the bicoloring ($q=2$) of $K=4$-uniform $\ell=19$-regular random hypergraphs. It is indeed for this connectivity that we found the coexistence of two solutions of the 1RSB equations at $m=1$ (and zero temperature), one with a high value of the overlap and a positive complexity, one with a low value of the overlap and a negative complexity, exactly as in the coexistence region of the ER ensemble. The results presented in Fig.~\ref{fig:regT01} show the evolution of these two solutions as $m$ is varied (we used here a small positive temperature $T=0.1$, but the $T=0$ case is qualitatively equivalent). We see that the coexistence phenomenon persists in an interval of $m$, roughly for $m \in [0.35,1]$, and that the HO branch disappears for smaller values of $m$. More importantly one sees that the complexity of the HO branch remains positive for $m \le 1$ (and even increases when $m$ decreases), while the complexity of the LO branch changes sign at the static value of the Parisi parameter $m_{\rm s}\approx 0.26$. This confirms the sketch of Fig.~\ref{fig:sketch_Phi}, and in particular the fact that the solution that minimizes $\Phi(m)/m$ is the continuation of the LO branch (or more generically of the one with the most negative complexity at $m=1$).

We also studied the same regular ensemble as a function of temperature, see the results in Fig.~\ref{fig:regm1}. Upon lowering the temperature a non-trivial solution of the 1RSB equation at $m=1$ appears continuously at the Kesten-Stigum threshold \eqref{eq:KSwithT} with a negative complexity; this solution persists down to zero temperature and corresponds to the LO branch. The other branch appears discontinuously, with a positive complexity, around $T=0.1$, and is always thermodynamically irrelevant according to our analysis. Lowering the temperature in the regular ensemble is thus qualitatively similar to increasing the average degree at zero temperature in the ER ensemble.

\subsection{Asymptotic expansions}
\label{sec:asymptotic}

We shall discuss now the asymptotic scalings of the various thresholds described in this paper when $K$ and/or $q$ gets large, the details of the computations being deferred to Appendix~\ref{app:asympt}.

The easiest one to begin with is the upperbound on the colorability threshold provided by the RS cavity method, or equivalently with the first moment method. As it is given explicitly (see \eqref{eq:lRS}) it can be expanded as 
\begin{align}
\ell_{\rm{RS}} =  K q^{K-1} \ln q - \frac{1}{2} K \ln q + O\left( \frac{K \ln q}{q^{K-1}}\right)\; .
\end{align}
Here the large parameter in which one expands is $q^{K-1}$, hence the expansion is valid when $K$ and/or $q$ gets large, as in all of these cases $q^{K-1} \to \infty$.

We also considered the rigidity tresholds $\ell_{\rm r}(m)$, defined as the smallest degree above which the clusters described by the 1RSB equations with Parisi parameter $m$ contain an extensive number of frozen variables, and found them to behave as
\begin{align}
\label{eq:main_asymptot_lr}
\ell_{\rm{r}}(m) = q^{K-1} [& \ln(K-1)+ \ln(q-1) \\ & + \ln\ln(Kq) + C(m) + o(1) ] \ , \nonumber
\end{align}
an expansion valid when $q$ and/or $K$ diverges (see Appendix~\ref{app:asympt} for details). The constant term in the parenthesis depends on the Parisi parameter $m$, we computed its value in the two special cases $m=1$ and $m=0$, finding $C(m=1)=1$ and $C(m=0)=1-\ln 2$ (in agreement with previous results for $K=2$~\cite{Krzakala2004,Zdeborova2007} and $q=2$~\cite{BrDaSeZd16}). We also expect the clustering transition $\ell_{\rm clust}$ to have the same asymptotic expansion as in \eqref{eq:main_asymptot_lr}, with a different constant term $C$; this has been rigorously proven for coloring of graphs ($K=2$) in~\cite{Sly08,Sly16}.

The colorability and condensation thresholds have been found in previous studies~\cite{Krzakala2004,Mertens2005,Zdeborova2007,MoRiSe08,BrDaSeZd16} to occur asymptotically very close to the upperbound of the first moment method. This is also the case here, we have indeed obtained:
\begin{align}
\ell_{\rm{col}} =  K q^{K-1} \ln q & - \frac{1}{2} K \left(\ln q + 1 - \frac{1}{q} \right) \label{eq:main_asymptot_lcol} \\ & 
+ \tO\left( \frac{1}{q^{K-1}}\right)\; , \nonumber \\
\ell_{\rm{cond}}  =  K q^{K-1} \ln q &- \frac{1}{2} K \left(\ln q + \left(1 - \frac{1}{q}\right) 2 \ln 2 \right) \label{eq:main_asymptot_lcond} \\ & 
+ \tO\left( \frac{1}{q^{K-1}}\right)\; .
\nonumber
\end{align}
To be more precise, we have performed these expansions by taking the limit $K \to \infty$ for a fixed value of $q$, the notation $\tO$ hiding terms which are polynomial in $K$, the expansion being naturally organized in inverse powers of the exponential scale $q^{K-1}$. However, setting $K=2$ in \eqref{eq:main_asymptot_lcol} and \eqref{eq:main_asymptot_lcond} we reproduce the results presented for the large $q$ limit of the $K=2$ graph case in~\cite{Krzakala2004,Zdeborova2007}, provided one removes the $1/q$ terms which are indeed comparable to the neglected terms of order $1/q^{K-1}=1/q$. We can thus conjecture these expansions to be valid when $K\ge 3$ and/or $q$ gets large, with $\tO$ hiding terms at most polynomial in $K$ and $\ln q$. These results are compatible with the asymptotic lowerbound on $\ell_{\rm col}$ presented in~\cite{Ayre2015} where the large $q$ limit is studied for arbitrary $K \ge 3$.

\section{Discussion}
\label{sec:conclu}
We summarize some of the main original features of the hypergraph coloring
problem and discuss their potential consequences:

\begin{itemize}
\item Survey Propagation, and therefore the 1RSB solution at $m=0$, is unstable
  around the colorability threshold for $q=2$, $K=3$ and $K=4$. While
  the instability of the 1RSB solution was observed in many models (the
  Sherrington-Kirkpatrick (SK) model and the satisfiability problem in the
  unsatisfiable phase 
  at low temperature~\cite{parisi1979infinite,crisanti20023} 
  being two notable examples), it is
  important to note that among previously studied random constraint
  satisfaction problems we are aware of only one case, the circular
  5-coloring of 3-regular graphs~\cite{schmidt2016circular}, where 1RSB has been
  found unstable at equilibrium in the satisfiable phase. The present case of $K=3$ or $4$
  hypergraph bicoloring is another example. In the other random CSPs where
the stability of SP has been tested, for instance random 3- and 4-SAT~\cite{MontanariParisi04}, instabilities were found in a part of the satisfiable phase that did not include the SP prediction of the satisfiability threshold.

Establishing this example
  is important for the following reason: while at least on the
  physics level the 1RSB picture is well understood in sparse
  constraint satisfaction problems, there is no known solvable form of
  full-step-replica-symmetry-breaking (FRSB) equations for sparse
  systems (see~\cite{Parisi2017} for a recent attempt in this direction). 
The nature of the structure of solutions in a CSP where
  1RSB is not stable is a widely open problem (even on the physics
  non-rigorous level). One can conjecture this structure to be similar
to the FRSB phase that has been well described
  and understood in dense models, such as the SK model
  \cite{MezardParisi87b,talagrand2003spin,Panchenko_book}, yet this remains to be put on
a more solid ground; having concrete examples on which to study this phenomenon is thus an important starting point. An important difference between FRSB in the satisfiable regime of CSPs and in the SK model is the fact that in the former case the groundstate energy is known to be exactly zero, while in the latter it is a non-trivial quantity. This fact might simplify the analysis, and is reminiscent of the situation in the mean-field treatment of the hard-sphere model~\cite{ChKuPaUrZa17} where a so-called Gardner transition towards FRSB occurs in the un-jammed (satisfiable) phase.

\item We found a coexistence of two 1RSB solutions in the bicoloring ($q=2$) of
  $K \ge 4$-uniform hypergraphs. We concluded
  that this coexistence has no significant bearings for the
  thermodynamics of random ensembles as studied in this
  paper. However, we note that this phenomenon
  is much more interesting in the perspective of planted inference
  problems, as discussed in detail in~\cite{RiSeZd17}.

\item We have obtained the asymptotic expansions of the rigidity, condensation and coloring thresholds when the parameters $q$ and/or $K$ diverge. These expansions connect smoothly (at least at the order of the expansion we have reached) the one previously obtained for $q=2$ and $K=2$.

\end{itemize}

Concerning other open directions for future work we want to note that
the version of hypergraph coloring where each constraint requires no
more than $\gamma$ variables to have the same color for 
$\gamma<K-1$~\cite{Krivelevich1998}
is also interesting and might also bring examples of interesting new
phenomena. 

Establishing rigorously the results presented here is
another natural direction for future work. 
In particular any result for the case of bicoloring
with $K=3$ and $K=4$ beyond the clustering threshold (equal to the
Kesten-Stigum bound in those cases) would be remarkable because it
might shed some light about the mysterious FRSB in sparse constraint
satisfaction problems. 

We also note that the freezing transition where all solutions acquire
frozen variables has been so far determined analytically only for the 
bicoloring of random hypergraph problem~\cite{BrDaSeZd16}. Generalizing the
results of~\cite{BrDaSeZd16} to $q >2$ should in principle be possible,
as a first step towards the computation of the freezing transition in
arbitrary random CSPs.

\begin{acknowledgments}

We warmly thank the Simons Institute for the Theory of Computing at UC Berkeley, where this work has been initiated as a working group within the program Counting Complexity and Phase Transitions, and in particular Florent Krzakala, Catherine Greenhill and Laura Florescu. M.G. acknowledges funding from ``Chaire de recherche sur les mod\`eles et sciences des donn\'es'', Fondation CFM pour la Recherche-ENS and from ERC grant SPARCS 307087.

\end{acknowledgments}

\bibliographystyle{unsrt}

\small{ \bibliography{references} }

\newpage
\appendix
\section{Threshold definitions}
\label{app:thresholds}
We summarize the definitions and notations for the various transitions undergone by random ensembles of CSPs in Table~\ref{tab:thresholds}.

\begin{table*}
\begin{center}
\begin{tabular}{p{0.12\linewidth}p{0.07\linewidth}p{0.7\linewidth}}
Transition / $\quad$ Instability & Notation & Definition / Cavity method evidence strategy / Other therminology\\
\hline \hline
Clustering    &$\ell_{\rm clust}$ 
        & The set of typical solutions splits into exponentially many clusters, at the threshold the MCMC equilibration time diverges. The 1RSB cavity equations at $m=1$ acquire a non-trivial solution.
\\
        &&{\footnotesize Other therminology: dynamical transition (in the context of mean field structural glasses), solvability of the tree reconstruction problem.} \\
\\
{RS - linear \newline stability}  & $\ell^{\rm{RS}}_{\rm{stab}}$
        & The trivial (RS) solution of the 1RSB equations becomes linearly unstable under small perturbations. This provides an upperbound on $\ell_{\rm clust}$. \\
        && {\footnotesize Other therminology: Kesten-Stigum transition for tree reconstruction, stability of the uninformative fixed point (for planted problems).  }\\
\\
Condensation  &$\ell_{\rm cond}$ 
        & A finite number of clusters come to dominate the set of solutions. The total entropy is not equal to the RS entropy anymore and becomes strictly smaller. The complexity (computed at $m=1$) becomes negative.  \\
        && {\footnotesize Other therminology: Kauzmann or ideal glass transition (mean field structural glasses), information-theoretic limit (for inference of planted models).} \\
\\
Rigidity    &$\ell_{\rm r}$
        & Typical solutions start to have a fraction (bounded away from zero) of frozen variables. The 1RSB population dynamics (at the static value of the Parisi parameter $m=m_{\rm s}$) features some hard fields. The notation $\ell_{\rm r}(m)$ indicates the transition for a prescribed value of $m$. It is much easier to compute it for $m=1$ or $m=0$ than for arbitrary values of $m$. If $\ell_{\rm r}(m=1) < \ell_{\rm cond}$ then $\ell_{\rm r}=\ell_{\rm r}(m=1)$; this happens in the generic scenario for $q,K$ sufficiently large. $\ell_{\rm r}(m=1)$ is an upperbound for $\ell_{\rm clust}$.
\\
        && {\footnotesize Other therminology: sometimes unhappily refered to as freezing; $\ell_{\rm r}(m=1)$ is the threshold for naive reconstruction.} \\
\\
SP          & $\ell^{\rm SP}$ 
        & The SP equation starts to have a non trivial solution. Corresponds to $\ell_{\rm r}(m=0)$.
\\
\\
Freezing    &$\ell_{\rm f}$   
        & All solutions start to have a fraction bounded away from zero of frozen variables. \\

\\
Colorability  &$\ell_{\rm col}$  
        & There exists no more valid colorings. The frozen variables start to break constraints. The complexity at $m=0$ becomes negative as all clusters disappear. \\
        && {\footnotesize Other therminology: satisfiability (generic CSPs)} \\
\\
RS            &$\ell_{\rm RS}$  
        & The RS entropy reaches zero. This threshold is an upper bound of the true colorability threshold. \\
        && {\footnotesize Other therminology:
                                   In problems with uniform BP fixed
                                   point this is the first moment bound} \\

\\
SP - type I \newline stability   & $\ell^{\rm SP}_{\rm I}$
              & Noise propagation: the SP non trivial fixed point is linearly unstable under small perturbations breaking the color symmetry. \\
\\          
SP - type II \newline stability  & $\ell^{\rm SP}_{\rm II}$
        & Bug proliferation: changing the color of one frozen variable implies changes at long range \\
\hline
\end{tabular}
\end{center}
\caption{Definitions, notation and principal characteristics of the main transitions undergone by the set of solutions of random CSPs. 
}
\label{tab:thresholds}
\end{table*}

\section{Stability of the RS solution}
\label{app:LinearStab_RS}
We shall prove here the Kesten-Stigum bound \eqref{eq:KSbound} on the clustering transition. To do so we shall study the stability of the trivial RS solution of the 1RSB equations, and show that for $\ell > \ell^{\rm{RS}}_{\rm{stab}}$ it is locally unstable. The iterations of the 1RSB equations must then flow to another fixed-point, which completes the reasoning as $\ell_{\rm clust}$ is defined in terms of the existence of a non-trivial solution of the 1RSB equations.

To put this reasoning on a quantitative basis let us consider a distribution $P(\psi)$ in the support of $\Pm^{\rm{1RSB}}(P)$, that is close to the RS distribution $\delta[\psi - \overline{\psi}]$. To quantify the distance between $P$ and this Dirac delta it is natural to consider its first moments,
\begin{align}
&M(x) = \int {\rm d} \psi P(\psi) \left(\psi(x) - \frac{1}{q}\right) \, , \\
&V(x,x') = \int {\rm d} \psi P(\psi) \left(\psi(x) - \frac{1}{q}\right)
\left(\psi(x') - \frac{1}{q}\right) \, .
\end{align}
We note however that the condition \eqref{eq:P_sym} that enforces the symmetry between colors in the 1RSB formalism implies that $M=0$. Hence at the lowest order the closeness of $P$ from the Dirac measure in $\overline{\psi}$ can be tracked by the covariance matrix $V(x,x')$. The latter has itself to respect the invariance under color permutations (hence can only depend on whether $x=x'$ or not), the symmetry $V(x,x')=V(x',x)$, and the normalization condition $\sum_x V(x,x')=0$, as the $\psi$ in the support of $P$ are probability distributions over $\mathcal{Q}$. Consequently $V$ must be of the form $V(x,x')=v (q\delta_{x,x'}-1)$, with $v$ a scalar quantity quantifying the width of $P$. What remains now to assess is the growth or shrinking of these $v$'s under the iterations of the 1RSB equation \eqref{eq:pdfmsgmsg}. 

To do so suppose that the distributions $\{P\itm\}$ are characterized by variances $\{v\itm\}$, and let us compute the variance $v$ of $P=\FRSB(\br{P\jtn})$. As a first step we consider the RS iteration $\psi=\FRS(\br{\psi\jtn})$, with arguments $\psi\jtn(x) = \frac{1}{q} + \epsilon\jtn (x) $ close to the uniform distribution $\overline{\psi}$. Expanding \eqref{eq:defZ0} linearly in the $\epsilon$'s yields
\begin{align}
\psi(x) = \frac{1}{q} + \frac{e^{-\beta} -1}{q^{K-1} + (e^{-\beta} -1)} 
\sum_{\nu=1}^d \sum_{j=1}^{K-1} \epsilon\jtn (x) \ ,
\end{align}
where we exploited the normalization condition $\sum_x \epsilon\jtn (x) =0$. Injecting this expansion in \eqref{eq:nd1RSB}, and noticing that the normalization $\Z_1$ and the reweighting factor $\Z_0^m$ compensates at lowest order in the variance expansion, we obtain a projection of $P=\FRSB(\br{P\jtn})$ in terms of the variances as
\begin{align}
v =  \left(\frac{e^{-\beta} -1}{q^{K-1} + (e^{-\beta} -1)} \right)^2
\sum_{\nu=1}^d \sum_{j=1}^{K-1} v\jtn  \ .
\end{align}
This relation being linear in the $v$'s one can take its average with respect to $\Pm^{\rm{1RSB}}$ and deduce that the instability of the trivial RS solution under the iterations of \eqref{eq:pdfmsgmsg} will occur if and only if
\begin{align}
\left(\sum_{d=1}^\infty r_d d\right) (K-1) \left(\frac{e^{-\beta} -1}{q^{K-1} + (e^{-\beta} -1)} \right)^2 > 1 \ .
\label{eq:KSwithT}
\end{align}
For a Poisson distribution of mean $\ell$, and at zero temperature, this gives the criterion stated in \eqref{eq:KSbound}.

\section{Survey propagation instabilities}
\label{app:SP}

In this appendix we give some details on the computation of the instability parameters $\lambda_{\rm I}$ and $\lambda_{\rm II}$ introduced in Section~\ref{subsec:col}. As explained there these parameters allow to test the existence of proper 2RSB solutions of the cavity equations that are close to the SP solution. As the 2RSB formalism describes the configuration space with two hierarchical levels in the organization of the pure states there are two ways to embed the SP description as a degenerate 2RSB solution: either there is only one group of pure states, or each pure state contains only one configuration. We shall hence study the perturbations of the SP equations in the neighborhood of these two reductions and assess their stability.

\subsection{Type I instability}

The stability analysis of SP under noise propagation (type I instability) is somewhat similar to the linear stability of the RS solution detailed in Appendix \ref{app:LinearStab_RS}. We shall indeed replace, on a given link $i \to \mu$ of the factor graph, the SP message $\eta\itm$ by a narrow distribution $P\itm$ of such messages, and check whether this distribution shrinks or expands upon iteration. More quantitatively we shall follow the average of $P\itm$ and its covariance, to be denoted $V\itm$. We assume that $P\itm$ is symmetric under the permutation of colors, hence its average is a vector with $q$ equal elements, while its covariance is a $q \times q$ matrix with only two distinct elements (on the diagonal and outside). 

To obtain the evolution equation of the covariance matrices we shall first write a generalization of the Survey Propagation equation \eqref{eq:SPrec}, without assuming the symmetry under permutation of the $q$ colors. Projecting \eqref{eq:nd1RSB} on the intensity of the hard fields one finds
\begin{align}
\label{eq:SPbroken}
\eta\itm^x = \frac{\underset{S \subset \mathcal{Q}\setminus x}{\sum}(-1)^{|S|} \underset{\nu \in \partial i \setminus \mu}{\prod} \left(1 - \xi\nti^x - \underset{x' \in S}{\sum} \xi\nti^{x'} \right) }
{\underset{S \subset \mathcal{Q}, |S| \ge 1}{\sum}(-1)^{|S|+1} \underset{\nu \in \partial i \setminus \mu}{\prod} \left(1 - \underset{x' \in S}{\sum} \xi\nti^{x'} \right)
}
\end{align}
where
\begin{align}
\label{eq:SPedmsg}
\xi\nti^x = \prod\nbe{j}{\nu}{i} \n\jtn^x \, ,
\end{align}
and we recall that $\mathcal{Q} = \{1,\dots,q\}$ denotes the set of available colors. 

A small perturbation of one incoming message $\eta\jtn^y$ in \eqref{eq:SPbroken} will induce a small perturbation of the outcoming message $\eta\itm^x$. To quantify this effect at linear order we define the $q\times q $ matrix $T\itm\jtnu$ by
\begin{align}
\s{T\itm\jtnu }_{x,y}= \left. \pd{\n\itm^x}{\n\jtn^y} \right|_{\rm sym} \, ,
\end{align}
where the subscript ``$\rm sym$'' means that we evaluate the derivative at the color-symmetric point with all the incoming and outcoming $\eta$'s independent of their color index. After a short computation one finds that the matrix elements of $T$ take only two distinct values, depending on whether $x=y$ or not. More precisely, the diagonal and off-diagonal elements of $T$ are, respectively:
\begin{widetext}
\begin{align}
\s{T\itm\jtnu}_{x,x}  &= 
  \p{\prod\nbe{k}{\nu}{i,j} \n\ktn} \p{1 - \n\itm} \; 
\frac{\displaystyle\sum_{p=0}^{q-1} (-1)^{p+1} \dbinom{q-1}{p} 
\prod\nbe{\rho}{i}{\mu,\nu}  \p{1 - \p{p+1} \prod\nbe{j}{\rho}{i} \n\jtr} }
{\displaystyle\sum_{p=1}^{q} (-1)^{p+1} \dbinom{q}{p} 
\prod\nbe{\rho}{i}{\mu}  \p{1 - p \prod\nbe{j}{\rho}{i} \n\jtr} }
\; ,  \\
\s{T\itm\jtnu}_{x,y} & =  \p{\prod\nbe{k}{\nu}{i,j} \n\ktn} \; 
\left( \; 
\frac{\displaystyle\sum_{p=0}^{q-2} (-1)^p \dbinom{q-2}{p} 
\prod\nbe{\rho}{i}{\mu,\nu}  \p{1 - \p{p+2} \prod\nbe{j}{\rho}{i} \n\jtr} }
{\displaystyle\sum_{p=1}^{q} (-1)^{p+1} \dbinom{q}{p} 
\prod\nbe{\rho}{i}{\mu}  \p{1 - p \prod\nbe{j}{\rho}{i} \n\jtr} } \right. \\
 & \qquad \qquad \qquad \qquad \qquad \qquad \qquad
 \left. - \;  \n\itm
\frac{\displaystyle\sum_{p=0}^{q-1} (-1)^{p+1} \dbinom{q-1}{p} 
\prod\nbe{\rho}{i}{\mu,\nu}  \p{1 - \p{p+1} \prod\nbe{j}{\rho}{i} \n\jtr} }
{\displaystyle\sum_{p=1}^{q} (-1)^{p+1} \dbinom{q}{p} 
\prod\nbe{\rho}{i}{\mu}  \p{1 - p \prod\nbe{j}{\rho}{i} \n\jtr} } \right)
\; ,
\end{align}
\end{widetext}
for all $x \neq y$. Such a matrix has two distinct eigenvalues, $T_{x,x} + (q-1) T_{x,y}$ with constant eigenvector, and $T_{x,x} - T_{x,y}$, $q-1$-times degenerate, with eigenspace perpendicular to the constant vector. In the following we shall only need the latter eigenvalue, that will be denoted $\theta\itm\jtnu$. One finds after a short computation that the above expressions of the matrix elements imply
\begin{widetext}
\begin{align}
\theta\itm\jtnu = - \p{\prod\nbe{k}{\nu}{i,j} \n\ktn}
\frac{\displaystyle\sum_{p=0}^{q-2} (-1)^{p} \dbinom{q-2}{p} 
\prod\nbe{\rho}{i}{\mu,\nu}  \p{1 - \p{p+1} \prod\nbe{j}{\rho}{i} \n\jtr} }
{\displaystyle\sum_{p=1}^{q} (-1)^{p+1} \dbinom{q}{p} 
\prod\nbe{\rho}{i}{\mu}  \p{1 - p \prod\nbe{j}{\nu}{i} \n\jtr} } \ .
\label{eq:theta_typeI}
\end{align}
\end{widetext}
We can indeed concentrate on the fluctuations of $\eta$ that are perpendicular to the constant vector: the longitudinal one does not bring out of the SP solution studied in \eqref{eq:pdfSPmsg}, an instability in this direction would be associated to a bifurcation of \eqref{eq:pdfSPmsg}. Hence $V$ can be parametrized by a single scalar $v$, as in Appendix \ref{app:LinearStab_RS}. Exploiting the computation made above on the derivative of $\FSP$ one sees that the joint distribution of the average and variance of the $P\itm$ evolve under iteration as
\begin{widetext}
\begin{align}
\Pm^{(n+1)}(\n,v) =  \sum_{d=0}^{\infty} r_d & \int \prod_{\nu = 1}^d \prod_{j=1}^{K-1} \mathrm{d} \n\jtn  \mathrm{d} v\jtn  
\Pm^{(n)}(\n\jtn,v\jtn) \ 
\kr{\n}{\FSP(\br{\n\jtn})} \kr{v}{\sum_{\nu=1}^d \sum_{j=1}^{K-1}  (\theta \jtnu)^2 v\jtn }
\;.
\label{eq:SP_stabI}
\end{align}
\end{widetext}
The SP solution is type I stable if under these iterations the $v$'s shrink to zero, unstable if they grow indefinitely. In practice one solves this distributional equation via a population dynamics algorithm, representing $\Pm^{(n)}(\n,v)$ by a sample of pairs $\{(\eta_i^{(n)},v_i^{(n)})\}$. After each iteration one divides the $v_i^{(n+1)}$ by a common constant $\lambda_{\rm I}^{(n)}$ in order to keep $\sum_i v_i^{(n)}$ independent of $n$ (note that the evolution of the $v$'s in \eqref{eq:SP_stabI} is linear, which allows such a rescaling). To increase the accuracy we define $\lambda_{\rm I}$ as the geometric mean of $\lambda_{\rm I}^{(n)}$ for many consecutive iterations, the solution is thus stable if and only if $\lambda_{\rm I} < 1$.

\subsection{Type II instability}

Let us now turn to the analysis of the second type of instability, and first explain its principle for a generic CSP. The zero-temperature BP equations for generic CSPs can be phrased in terms of ``warnings'' sent by variables to neighboring constraints, and vice versa. We denote $\chi$ the set of possible warnings (that depends on the specific form of the CSP), with the special value $0$ encoding the absence of warning, and consider the situation where one variable node has $n$ variable nodes at distance 1 in the cavity graph. The projection of the BP equations on the warnings gives rise to two functions: $\mathcal{V}(x_1,\dots,x_n)$ which is the indicator function of the event ``the $n$ warnings are not contradictory'', i.e. they allow at least one configuration for the variable considered, and $h(x_1,\dots,x_n)$, which is the value of the warning sent by this variable node (which is well-defined if and only if $\mathcal{V}(x_1,\dots,x_n)=1$). The SP equation relates probability distributions on $\chi$: suppose that the warning $x_i$ is emitted with probability $\eta_i^{x_i}$, independently for each of the neighbors. Then the probability law $\eta^x$ of the output warning, conditional on the absence of contradiction between in-coming warnings, is:
\begin{align}
\eta^x &= \frac{1}{\Z_{\rm SP}} \sum_{x_1,\dots,x_n} \eta_1^{x_1} \dots \eta_n^{x_n} \delta_{x,h(x_1,\dots,x_n)} \mathcal{V}(x_1,\dots,x_n) \ ,  \nonumber \\
\Z_{\rm SP}&= \sum_{x_1,\dots,x_n} \eta_1^{x_1} \dots \eta_n^{x_n} \mathcal{V}(x_1,\dots,x_n) \ .\label{eq:SPgeneric}
\end{align}
As a first step towards the type II instability analysis we consider a random process in which a pair $(x_i,x'_i)$ is emitted by each neighbor, with a joint law denoted $\eta_i^{x_i,x'_i}$; the joint law of the pair of outcoming messages, again conditioning on the absence of contradictions in both copies of the process, reads
\begin{widetext}
\begin{align}
\eta^{x,x'} = \frac{\underset{\substack{ x_1,\dots,x_n \\ x'_1,\dots,x'_n }}{\sum} \eta_1^{x_1,x'_1} \dots \eta_n^{x_n,x'_n} \delta_{x,h(x_1,\dots,x_n)} \mathcal{V}(x_1,\dots,x_n)\delta_{x',h(x'_1,\dots,x'_n)} \mathcal{V}(x'_1,\dots,x'_n)}{
\underset{\substack{ x_1,\dots,x_n \\ x'_1,\dots,x'_n }}{\sum} \eta_1^{x_1,x'_1} \dots \eta_n^{x_n,x'_n} \mathcal{V}(x_1,\dots,x_n) \mathcal{V}(x'_1,\dots,x'_n)
} \ .
\label{eq:SPjoint}
\end{align}
\end{widetext}
It is obvious that if the two copies of the incoming messages are strictly identical, i.e. if $\eta_i^{x_i,x_i} = \eta_i^{x_i} \delta_{x_i,x'_i}$ for all $i$, then this is also the case for the outcoming pair of warnings, with a probability $\eta^x$ given by \eqref{eq:SPgeneric}. The type II instability corresponds to a deviation from this strict coupling: we assume that $\eta_i^{x_i,x'_i}$ is very small for $x_i\neq x'_i$, and denote this quantity $\varepsilon_i^{x_i \to x'_i}$, while for $x_i = x'_i$ we keep the notation $\eta_i^{x_i}$. Expanding at first order in the $\varepsilon_i$ the equation \eqref{eq:SPjoint} one finds that the output joint law is also close to diagonal, with the small probabilities $\varepsilon^{x \to x'}$ given for $x \neq x'$ by
\begin{widetext}
\begin{align}
\varepsilon^{x \to x'} = \frac{1}{\Z_{\rm SP}} \sum_{i=1}^n \sum_{\substack{x_1,\dots,x_n \\ x'_i \neq x_i}}
\left(\prod_{j\neq i} \eta_j^{x_j}\right) \varepsilon_i^{x_i \to x'_i}  \delta_{x,h(x_1,\dots,x_n)} \mathcal{V}(x_1,\dots,x_n)\delta_{x',h(x_1,\dots,x'_i,\dots, x_n)} \mathcal{V}(x_1,\dots,x'_i,\dots, x_n) \ .
\end{align}
\end{widetext}
The name ``bug proliferation'' of this instability comes from the following algorithmic interpretation: the first copy of the process $x=h(x_1,\dots,x_n)$ is considered to be the ``correct'' one, while the primed warnings are corrupted by bugs which occurs with a small probability; the SP solution is unstable if the probabilities of the bugs grows during the propagation of the SP equations along the factor graph.

We shall now come back to the hypergraph coloring problem, and recalls that here the warning alphabet is $\{1,\dots,q,0\}$, the first $q$ values corresponding to a variable being forced to a given colour, while $0$ is sent if at least two colours are allowed for this variable. The warnings are propagated first from the variables to the hyperedges, a color being forbidden by an hyperedge if its $K-1$ neighbors are forced to the same color, then from hyperedges to variables, a variable being forced to a given value if the $q-1$ other colors are forbidden (a contradiction arises if the $q$ colors are forbidden). A change in a single warning can be propagated in the first step if the $K-2$ other warnings are forcing to the same color, and in the second step if $q-2$ colors are forbidden by the hyperedges not affected by the bug. One thus finds the following rules for the propagation of the bug probabilities,
\begin{widetext}
\begin{align}
\varepsilon\itm^{x \to 0} &= \frac{1}{\mathcal{Z}_{\rm SP}\itmu} \sum_{\substack{\nu \in \partial i \setminus \mu \\ j \in \partial \nu \setminus i}} \sum_{y \neq x}
\left( \sum_{p=0}^{q-2} (-1)^{p} \dbinom{q-2}{p} 
\prod\nbe{\rho}{i}{\mu,\nu}  \p{1 - \p{p+2} \prod\nbe{j}{\rho}{i} \n\jtr} \right) \left( \varepsilon\jtn^{y \to 0} + \sum_{y' \neq y} \varepsilon\jtn^{y \to y'} \right) \prod\nbe{k}{\nu}{i,j} \n\ktn \ , \nonumber \\ 
\varepsilon\itm^{0 \to x} &= \frac{1}{\mathcal{Z}_{\rm SP}\itmu}\sum_{\substack{\nu \in \partial i \setminus \mu \\ j \in \partial \nu \setminus i}} \sum_{y \neq x}
\left( \sum_{p=0}^{q-2} (-1)^{p} \dbinom{q-2}{p} 
\prod\nbe{\rho}{i}{\mu,\nu}  \p{1 - \p{p+2} \prod\nbe{j}{\rho}{i} \n\jtr} \right) \left( \varepsilon\jtn^{0 \to y} + \sum_{y' \neq y} \varepsilon\jtn^{y' \to y} \right) \prod\nbe{k}{\nu}{i,j} \n\ktn \ , \nonumber \\ 
\varepsilon\itm^{x \to x'} &= 0 \ , \nonumber
\end{align}
\end{widetext}
where $x,x',y$ and $y'$ denote non-trivial warnings (different from 0). Note that in the last line we assumed $K \ge 3$ (but the final result given below turns out to be valid also for $K=2$), and that in the first two lines the factor containing the summation over $p$ is the probability that the constraints not affected by the bug forbids the $q-2$ colors distinct from $x$ and $y$. After one iteration one has $\varepsilon^{x \to x'} =0$, and the $\varepsilon^{x \to 0}, \varepsilon^{0 \to x}$ become independent of colors. One can thus simplify these equations and keep a single $\varepsilon$ on each edge, that evolves according to
\begin{align}
\varepsilon\itm = (q-1) \sum_{\substack{\nu \in \partial i \setminus \mu \\ j \in \partial \nu \setminus i}} \mathcal{A}\itm\jtnu \varepsilon\jtn
\end{align}
where the $(q-1)$ comes from the summation over $y\neq x$ in the above equations, and where we defined
\begin{widetext}
\begin{align}
\mathcal{A}\itm\jtnu =  \p{\prod\nbe{k}{\nu}{i,j} \n\ktn}
\frac{\displaystyle\sum_{p=0}^{q-2} (-1)^{p} \dbinom{q-2}{p} 
\prod\nbe{\rho}{i}{\mu,\nu}  \p{1 - \p{p+2} \prod\nbe{j}{\rho}{i} \n\jtr} }
{\displaystyle\sum_{p=1}^{q} (-1)^{p+1} \dbinom{q}{p} 
\prod\nbe{\rho}{i}{\mu}  \p{1 - p \prod\nbe{j}{\nu}{i} \n\jtr} } \ .
\end{align}
\end{widetext}
This quantity is very similar to the eigenvalue $\theta\itm\jtnu$ defined in \eqref{eq:theta_typeI} during the study of the type I inequality, except for a global sign and a shift from $p+1$ to $p+2$ in the bracket of the numerator.

Finally the instability of typical random hypergraphs with excess degree distribution $r_d$ is studied through a population of couples $(\n,\varepsilon)$, evolving according to
\begin{widetext}
\begin{align}
\Pm^{(n+1)}(\n,\varepsilon) =  \sum_{d=0}^{\infty} r_d & \int \prod_{\nu = 1}^d \prod_{j=1}^{K-1} \mathrm{d} \n\jtn  \mathrm{d} \varepsilon \jtn  
\Pm^{(n)}(\n\jtn,\varepsilon\jtn) \ 
\kr{\n}{\FSP(\br{\n\jtn})} \kr{\varepsilon}{\sum_{\nu=1}^d \sum_{j=1}^{K-1}  (q-1) \mathcal{A}\jtnu \varepsilon \jtn }
\; ,
\label{eq:SP_stabII}
\end{align}
\end{widetext}
an equation similar to \eqref{eq:SP_stabI} used for the study of the type I instability. As explained in that case we define $\lambda_{\rm II}$ as the quantity by which the $\varepsilon$ should be divided to keep their norm constant under these iterations.

\section{Details on the asymptotic expansions}
\label{app:asympt}
In this section we present some more details on the computations that yield the asymptotic expansion of the thresholds presented in \ref{sec:asymptotic}. In these limits of large $K$ and/or $q$ the relevant average degree diverges, the Poisson random graph ensemble with average degree $\ell \gg 1$ and the regular ensemble with degree $\ell \gg 1$ can be considered as equivalent at leading order as the Poisson distribution concentrates around its average value. Depending on the threshold to be determined we shall thus use the most convenient ensemble between these two.

\subsection{Rigidity threshold at $m=1$}

Consider the equation \eqref{eq:recursion_rigidity_m1} that gives the probability of a frozen variable in the 1RSB cavity formalism at $m=1$, for a Poissonian degree distribution with mean $\ell$. We rewrite it as
\begin{equation}\label{eqn:etaER}
x =f(x,\ell)= \left(1- e^{-\ell \frac{x^{K-1}}{q^{K-1}-1}}\right)^{q-1} \ ,
\end{equation}
where we defined for simplicity $x=q \eta$. The rigidity threshold $\ell_{\rm r}$ is defined as the smallest value of $\ell$ such that there exists a strictly positive solution $x_{\rm r}$ of this fixed point equation. As $f$ behaves as $x^{(K-1)(q-1)}$ around $x=0$, the bifurcation at $\ell_{\rm r}$ has to occur discontinuously whenever $(K-1)(q-1)>1$, a condition that we assume in the following, in which we shall denote $L=(K-1)(q-1)$. Determining $\ell_{\rm r}$ thus amounts to solve for $\ell_{\rm r}$ and $x_{\rm r}>0$ solution of
\begin{align}
\begin{cases}
x_{\rm r} =f(x_{\rm r},\ell_{\rm r}) \\
1 =\left. \frac{\partial f}{\partial x} \right|_{(x_{\rm r},\ell_{\rm r})}
\end{cases} \ .
\end{align}
These two equations are easily shown to be equivalent to
\begin{align}
\label{eq:asymptlr}
\begin{cases}
\ell_{\rm r} = (q^{K-1} -1)\frac{1}{x_{\rm r}^{K-1}} \ln\left(\frac{1}{1-x_{\rm r}^{1/(q-1)}}\right) 
\\
1 = L (1-x_{\rm r}^{-1/(q-1)}) \ln (1-x_{\rm r}^{1/(q-1)})
\end{cases} \ .
\end{align}
It now remains to solve for $x_{\rm r}$ in the second equation and replace in the first one to obtain the asymptotic expansion of $\ell_{\rm r}$. Denoting $y=1/(1-x_{\rm r}^{1/(q-1)})$, the second equation fixes the value of $y(L)$ according to
\begin{align}
1 = L \frac{1}{y(L)} \frac{1}{1-\frac{1}{y(L)}} \ln y(L) \ .
\end{align}
To obtain the asymptotic expansion of $y(L)$ when $L \to \infty$ we rewrite this as
\begin{eqnarray}
y(L) &=& L \ln y(L)  \frac{1}{1-\frac{1}{y(L)}} \\
&=& L \left(\ln L + \ln \ln y(L) - \ln\left(1-\frac{1}{y(L)}\right)  \right) \frac{1}{1-\frac{1}{y(L)}} \nonumber
\end{eqnarray}
From this equation one easily sees that $y(L) \sim L \ln L$, and then more precisely that
\begin{eqnarray}
y(L) &=& L \left( \ln L + \ln\ln L + O\left( \frac{\ln \ln L}{\ln L}\right) \right) \ , \\
\ln y(L) &=& \ln L + \ln\ln L + O\left( \frac{\ln \ln L}{\ln L}\right) \ .
\end{eqnarray}
Coming back to the determination of the asymptotic behavior of $\ell_{\rm r}$ we note that
\begin{eqnarray}
\frac{1}{x_{\rm r}^{K-1}} \ln\left(\frac{1}{1-x_{\rm r}^{1/(q-1)}}\right) &=& \ln (y(L)) \left(1 - \frac{1}{y(L)} \right)^{-L} \\ &=& \ln L + \ln\ln L + 1 + O\left( \frac{\ln \ln L}{\ln L}\right) 
\nonumber
\end{eqnarray}
Inserting this expansion in the expression of $\ell_{\rm r}$ given in \eqref{eq:asymptlr} yields the formula we stated in \eqref{eq:main_asymptot_lr}, noting finally that $\ln\ln L = \ln\ln (Kq) + o(1)$ as soon as $L$ diverges, whenever $K$ and/or $q$ go to infinity.

\subsection{Rigidity threshold at $m=0$}

We turn now to the determination of the rigidity treshold at $m=0$, in other words the smallest degree for which the SP equation \eqref{eq:pdfSPmsg} admits a non-trivial solution. This computation turns out to be easier in the regular ensemble, for which $\Pm^{\rm{SP}}$ is a Dirac distribution on $\eta$ solution of
\begin{align}
\label{eq:SPreg}
\eta= \frac{\underset{p=0}{\overset{q-1}{\sum}} (-1)^p\binom{q-1}{p} (1-(p+1)\eta^{K-1})^{\ell-1}}{\underset{p=1}{\overset{q}{\sum}} (-1)^{p+1}\binom{q}{p} (1-p\eta^{K-1})^{\ell-1}} \ ,
\end{align}
as follows from \eqref{eq:SPrec}. Inspired by the study of the $m=1$ rigidity we consider the following scale of degrees,
\begin{align}
\ell = q^{K-1}(\ln L + \ln\ln L + \ell'+o(1)) \, ,
\end{align}
with $\ell'$ a constant, and assume that the solution $\eta$ of \eqref{eq:SPreg} behaves in this case as
\begin{align}
\eta = \frac{1}{q} \left(1- \frac{\eta'}{(K-1) \ln L} + o\left(\frac{1}{(K-1) \ln L}  \right)\right) \ ,
\end{align}
where $\eta'$ is a constant to be determined as a function of $\ell'$. To do so we first note that as $L\to\infty$ one has
\begin{align}
\eta^{K-1} = \frac{1}{q^{K-1}} \left(1- \frac{\eta'}{\ln L} + o\left(\frac{1}{\ln L}  \right)\right) \ .
\end{align}
Then we compute
\begin{align}
(1-p\eta^{K-1})^{\ell-1} & = \exp\left[-p (\ln L + \ln\ln L + \ell'+o(1)) \right.
\nonumber \\
& \hspace{1.5cm} \left. 
\left(1- \frac{\eta'}{\ln L} + o\left(\frac{1}{\ln L}  \right)\right) \right] 
\nonumber \\
& \sim \left(\frac{1}{L\ln L} e^{\eta'-\ell'} \right)^p \ .
\label{eq:def_u}
\end{align}
Finally we note that
\begin{align}
\frac{\underset{p=0}{\overset{q-1}{\sum}} (-1)^p\binom{q-1}{p} u^{p+1}}{\underset{p=1}{\overset{q}{\sum}} (-1)^{p+1}\binom{q}{p} u^p} &= \frac{u(1-u)^{q-1}}{1-(1-u)^q} \\
&= \frac{1}{q}\left(1-\frac{q-1}{2} u + O((qu)^2) \right) \ , \nonumber 
\end{align}
and apply this identity with $u$ the quantity defined in \eqref{eq:def_u}. This gives us, from \eqref{eq:SPreg} and the ansatz made on $\eta$,
\begin{align}
1- \frac{\eta'}{(K-1) \ln L} \sim 1 - \frac{q-1}{2} \frac{1}{L\ln L} e^{\eta'-\ell'} \ .
\end{align}
Recalling that $L=(K-1)(q-1)$ we see that the ansatz was indeed self-consistent, and that $\eta'$ is determined as a function of $\ell'$ through
\begin{align}
\eta' e^{-\eta'} = \frac{1}{2} e^{-\ell'} \ .
\end{align}
This equation admits solutions only if $\ell' > 1- \ln 2$, hence the statement on $C(m=0)$ made in the main text.

\subsection{Colorability threshold}

To compute the asymptotics of the colorability transition we shall also work in the regular ensemble, we thus have to find $\eta$ solution of \eqref{eq:SPreg}, compute the complexity
\begin{align}
\Sigma(m=0)= & \ln\left(\sum_{p=1}^{q} (-1)^{p+1} \binom{q}{p} (1-p\eta^{K-1})^\ell\right) \nonumber \\ & - \frac{\ell (K-1)}{K} \ln\left(1-q \eta^K \right) \ ,
\label{eq:complexitySPreg}
\end{align}
and determine the $\ell$ where it vanishes.

We shall fix some $q \ge 2$ and take the limit $K \to \infty$; this allows to organize the asymptotic expansions with exponential dependency on $K$ dominating the polynomial ones. We shall in particular consider degrees on the scale 
\begin{align}
\ell(K)=K q^{K-1} \ln q - \ell_1(K) &- \frac{1}{q^{K-1}} \ell_2(K) 
\label{eq:scale_lcol} \\
& + \tO\left( \frac{1}{(q^{K-1})^2}\right)
\nonumber
\end{align}
with the $\ell_i$ polynomial in $K$, as we know that the colorability threshold coincides, at its leading order, with the upperbound given by the first moment method. We recall that the notation $\tO$ hides polynomial terms in $K$. Here we shall content ourselves with the first correction $\ell_1$, but it is not too difficult to generalize the computation to higher orders. For this scale of degrees we make the following ansatz on the solution $\eta$ of \eqref{eq:SPreg},
\begin{align}
\eta = \frac{1}{q} \left(1 - \frac{1}{q^{K-1}} \eta_1 - \tO\left( \frac{1}{(q^{K-1})^2}\right)  \right)
\label{eq:eta_asymptot}
\end{align}
with $\eta_1 = \tO(1)$. One finds after a short computation that
\begin{align}
(1- & p\eta^{K-1})^{\ell-1} =  \left(\frac{1}{q^K} \right)^p \\ & \left(1 + \frac{1}{q^{K-1}} \left[ p (\ell_1+1+K(K-1) \eta_1 \ln q) - \frac{1}{2} p^2 K \ln q \right] \right. \nonumber\\ & 
\left. + \tO\left( \frac{1}{(q^{K-1})^2}\right) \right) \nonumber
\end{align}
Plugging this result in \eqref{eq:SPreg} we can keep only the terms $p=0,1$ in the numerator, and $p=1,2$ in the denominator. This shows that the assumption \eqref{eq:eta_asymptot} is indeed self-consistent and fixes the coefficient $\eta_1 = (q-1)/(2q)$.
Performing the same kind of expansion on the expression \eqref{eq:complexitySPreg} of the complexity one finds
\begin{align}
\Sigma(m=0) = \frac{1}{q^{K-1}} \left[ \frac{\ell_1}{K} - \frac{1}{2} \left(
1 - \frac{1}{q} + \ln q \right)
\right] + \tO\left( \frac{1}{(q^{K-1})^2}\right) \ .
\end{align}
It is then trivial to deduce the value of $\ell_1$ for which the complexity vanishes at this order, this yields our asymptotic expansion for $\ell_{\rm col}$ stated in \eqref{eq:main_asymptot_lcol}.

\subsection{Condensation threshold}

We finally explain our computation of the asymptotic expansion for the condensation threshold given in \eqref{eq:main_asymptot_lcond}. We shall work in the Poissonian ensemble, with average degrees on the scale \eqref{eq:scale_lcol}, and determine the value of $\ell_1$ such that the complexity $\Sigma(m=1)=\Phi(m=1)-\Phi'(m=1)$ vanishes.

We can first easily expand $\Phi(m=1)=\phi_{\rm RS}$ on this scale to obtain from \eqref{eq:defRSent}:
\begin{align}
\Phi(m=1)= \frac{1}{q^{K-1}} \left[ \frac{\ell_1}{K} - \frac{1}{2} \ln q \right] 
+ \tO\left( \frac{1}{(q^{K-1})^2}\right) \ .
\end{align}
The non-trivial part of the computation is the expansion of $\Phi'(m=1)$. We shall use its expression in terms of the distribution $\overline{P}$ defined in \eqref{eq:def_overlineP},
\begin{align}
\Phi'(m=1) =& \sum_{d=0}^\infty p_d \int \prod_{\mu=1}^d \prod_{i=1}^{K-1} \mathrm{d} \psi\itm \overline{P}(\psi\itm) 
\frac{\Z_0^{i+\partial i}}{\overline{\Z}_0^{i+\partial i}} \ln \Z_0^{i+\partial i} \nonumber \\
&- \frac{\ell(K-1)}{K} \int \prod_{i=1}^K \mathrm{d} \psi_i \overline{P}(\psi_i) \frac{\Z_0^\mu}{\overline{\Z}_0^\mu} \ln \Z_0^\mu
\; ,
\label{eq:Phidevm1}
\end{align}
where $p_d$ is the Poissonian distribution of average $\ell$, and
\begin{align}
\Z_0^{i+\partial i} &= \sum_{x=1}^q \prod_{\mu=1}^d\left(1- \prod_{i=1}^{K-1}\psi\itm(x) \right)  \ , \\
\overline{\Z}_0^{i+\partial i} &= q \left(1 - \frac{1}{q^{K-1}} \right)^d \ , \\
\Z_0^\mu &= 1-\sum_{x=1}^q \prod_{i=1}^K \psi_i(x) \ , \\
\overline{\Z}_0^\mu &= 1 - \frac{1}{q^{K-1}} \ .
\end{align}
The distribution $\overline{P}$ is the solution of \eqref{eq:1RSBm1} for the excess law $r_d=p_d$. We decompose as usual the contribution of the hard fields with
\begin{align}
\overline{P}(\psi) = \eta \sum_{x=1}^q \delta[\psi - \psi^x] + (1- q \eta) \widetilde{P}(\psi) \ .
\end{align}
The weight $\eta$ of the hard fields is solution of \eqref{eq:recursion_rigidity_m1}; on this scale of degrees we let the reader check that it has the asymptotic expansion \eqref{eq:eta_asymptot}, with $\eta_1 = (q-1)/q$ (notice the factor 2 difference with the expansion at $m=0$). The fraction of soft fields is thus exponentially small in this regime; moreover one can safely assume that the soft fields are perfectly unbiased, i.e. that $\widetilde{P}(\psi) = \delta[\psi - \overline{\psi}]$. This is an approximation that does not change the estimate of the condensation threshold at this order.

Let us first consider the second line in \eqref{eq:Phidevm1}. We denote $K_0$, $K_1$, \dots, $K_q$ the number of the random $\psi_i$ which are equal, respectively, to $\overline{\psi}$, $\psi^1$, \dots, $\psi^q$. The distribution of $(K_0,K_1,\dots,K_q)$ is thus multinomial with parameters $(K;1-q\eta,\eta,\dots,\eta)$. One can easily compute the value of $\Z_0^\mu$ as a function of $(K_0,K_1,\dots,K_q)$; one finds that it is equal to 1 in many cases, except when $K_0=K$ (hence $K_1=\dots=K_q=0$) where one has $\Z_0^\mu = 1 - \frac{1}{q^{K-1}}$, or when there is exactly one color $x$ with $K_x >0$, then $\Z_0^\mu = 1 - \frac{1}{q^{K-K_x}}$. We can thus write
\begin{widetext}
\begin{align}
\int 
\prod_{i=1}^K \mathrm{d} \psi_i \overline{P}(\psi_i) \Z_0^\mu \ln \Z_0^\mu
& = 
(1-q \eta)^K \left(1- \frac{1}{q^{K-1}} \right) \ln\left(1- \frac{1}{q^{K-1}} \right) \nonumber \\ & 
+ q \sum_{K_x=1}^{K-1} \binom{K}{K_x} \eta^{K_x} (1-q \eta)^{K-K_x}
\left(1 - \frac{1}{q^{K-K_x}} \right) \ln\left(1 - \frac{1}{q^{K-K_x}} \right) \ ,
\end{align}
\end{widetext}
the only approximation made up to now being the replacement of $\widetilde{P}$ by $\delta[\psi - \overline{\psi}]$. We can then use the scaling \eqref{eq:eta_asymptot} of $\eta$: it implies that the leading behavior of this expression comes from the term $K_x=K-1$ in the sum, which gives
\begin{align}
&- \frac{\ell(K-1)}{K} \int \prod_{i=1}^K \mathrm{d} \psi_i \overline{P}(\psi_i) \frac{\Z_0^\mu}{\overline{\Z}_0^\mu} \ln \Z_0^\mu = \label{eq:Phideriv_clause} \\
& - \frac{1}{q^{K-1}} K (K-1) \frac{(q-1)^2}{q} (\ln q)\ln\left(1-\frac{1}{q}\right)
+ \tO\left( \frac{1}{(q^{K-1})^2}\right) \ . \nonumber
\end{align}
The first line of \eqref{eq:Phidevm1} can be handled in a similar way. We classify the $d$ constraints in different types according to the type of messages they receive from their $K-1$ variables, writing
\begin{align}
d = d_{\rm m} + d_{\rm s} + \sum_{x=1}^q \sum_{K_0=0}^{K-2} d_{x,K_0} \ .
\end{align}
In this expression $d_{\rm s}$ is the number of clauses that receive $K-1$ soft messages, $d_{x,K_0}$ counts those that receive $K-1-K_0$ messages of color $x$ and $K_0$ soft messages, and $d_{\rm m}$ counts all the other situations (i.e. those clauses that receive hard fields of at least two different colors). As $d$ is a random Poisson variable and as the type of each constraint is drawn independently of the others under the integrals in the first line of \eqref{eq:Phidevm1}, we see that these random degrees $\{d_0,d_{\rm s},\{d_{x,K_0}\} \}$ are independent Poisson variables, with averages
\begin{align}
\mathbb{E}[d_{\rm m}] = \ell p_{\rm m} \ , \ 
\mathbb{E}[d_{\rm s}] = \ell p_{\rm s} \ , \
\mathbb{E}[d_{x,K_0}] = \ell p_{K_0} \ .
\end{align}
The $p$'s are here the probabilities of the state of one constraint, from their definition one sees that
\begin{align}
p_{\rm s} &= (1-q \eta)^{K-1} \ , \\ 
p_{K_0} & = \binom{K-1}{K_0} \eta^{K-1-K_0} (1-q \eta)^{K_0} \ , \\ 
p_{\rm m} & = 1 - p_{\rm s} - q \sum_{K_0=0}^{K-2} p_{K_0} \ .
\end{align}
For a given state of this decomposition one can express $\Z_0^{i+\partial i}$  as
\begin{align}
\Z_0^{i+\partial i} = \left(1 - \frac{1}{q^{K-1}} \right)^{d_{\rm s}} 
\sum_{x=1}^q \mathbb{I}(d_{x,0}=0) \prod_{K_0=1}^{K-2}
\left(1 - \frac{1}{q^{K_0}} \right)^{d_{x,K_0}} \, .
\nonumber
\end{align}
Indeed the edges counted in $d_{x,0}$ have $K-1$ vertices forced in the color $x$, hence if $d_{x,0}>0$ the central vertex cannot take this color. On the other hand if at least two vertices are forced in at least two distinct colors the central vertex can take any color indifferently, hence $d_{\rm m}$ does not appear in this expression. The intermediate situations arise when exactly one color is forced in some of the $K-1$ neighbors in one constraint: the most numerous these are, the more the central vertex is biased to avoid this color.

It remains to compute the average of $(\Z_0 \ln \Z_0)/\overline{\Z}_0$ with the Poisson distribution of the various $d$. We first notice that on this asymptotic scale of degrees $\overline{\Z}_0 = (1/q^{K-1}) (1+ \widetilde{O}(1/q^{K-1}))$ and that the fluctuations in the Poisson degree $d$ can be neglected at the leading order, we can thus take $1/\overline{\Z}_0$ out of the average. Then we isolate the number $n$ of allowed colors for the central spin and write
\begin{align}
&\mathbb{E}[\Z_0^{i+\partial i} \ln \Z_0^{i+\partial i}] = \sum_{n=0}^q \binom{q}{n}
\left(e^{-\ell p_0}\right)^n \left(1-e^{-\ell p_0}\right)^{q-n}
 \nonumber \\ & \hspace{4cm} \mathbb{E}[\Z(n) \ln \Z(n)] \label{eq:Z0withn} \\
& \Z(n) = \left(1 - \frac{1}{q^{K-1}} \right)^{d_{\rm s}} 
\sum_{x=1}^n \prod_{K_0=1}^{K-2}
\left(1 - \frac{1}{q^{K_0}} \right)^{d_{x,K_0}} \nonumber
\end{align}
At this point we use the asymptotic behavior of $\eta$ and realize that
\begin{align}
e^{- \ell p_0} &= \frac{1}{q} \frac{1}{q^{K-1}} + 
\widetilde{O}\left(\frac{1}{(q^{K-1})^2}\right) \\
\ell p_1 &= K(K-1)(q-1) \ln q \frac{1}{q^{K-1}} + 
\widetilde{O}\left(\frac{1}{(q^{K-1})^2}\right) \\
\ell p_r &= \widetilde{O}\left(\frac{1}{(q^{K-1})^r}\right) \ \text{for} \ r \ge 2 \ ,
\end{align}
while $\ell p_{\rm s}$ is neglectible at all perturbative orders. The leading orders of \eqref{eq:Z0withn} thus comes from the smallest possible values of $n$. However $\Z(n=0)=0$ does not contribute to the sum, we have thus to consider $n=1$ and $n=2$. In the first case the most probable configuration with $d_{\rm s}=d_{1,1}=\dots=d_{1,K-2}=0$ will not contribute: this would yield $\Z=1$, hence cancelling the logarithm. The next-to-most probable configuration of the random variables is $d_{1,1}=1$, which happens with probability $\ell p_1 e^{-\ell p_1}$. From these observations we thus obtain
\begin{widetext}
\begin{align}
\mathbb{E}\left[\frac{\Z_0^{i+\partial i}}{\overline{\Z}_0^{i+\partial i}}  
\ln \Z_0^{i+\partial i}\right] & = q^{K-1} \left[ q \, e^{-\ell p_0} \ell p_1 \left(1-\frac{1}{q} \right) \ln \left(1-\frac{1}{q} \right) + \frac{q(q-1)}{2} \left(e^{-\ell p_0}\right)^2 2 \ln 2 
\right] \tO\left( \frac{1}{(q^{K-1})^2}\right) \\
& = \frac{1}{q^{K-1}} \left[ K (K-1) \frac{(q-1)^2}{q} (\ln q)\ln\left(1-\frac{1}{q}\right) + \left(1 - \frac{1}{q} \right) \ln 2 \right]
+ \tO\left( \frac{1}{(q^{K-1})^2}\right)
\end{align}
\end{widetext}
Adding up the contribution from \eqref{eq:Phideriv_clause} this yields
\begin{align}
\Phi'(m=1)= \frac{1}{q^{K-1}} \left(1-\frac{1}{q}\right) \ln 2 
+ \tO\left( \frac{1}{(q^{K-1})^2}\right) \ ,
\end{align}
hence the leading order of the complexity is
\begin{align}
\Sigma(m=1)&=\frac{1}{q^{K-1}} \left[ \frac{\ell_1}{K} - \frac{1}{2} \ln q - 
\left(1-\frac{1}{q}\right) \ln 2 \right] \nonumber \\
&+ \tO\left( \frac{1}{(q^{K-1})^2}\right) \ , \nonumber
\end{align}
which completes our justification of \eqref{eq:main_asymptot_lcond}.

\end{document}